\documentclass[aps,amsmath,amssymb,amsfonts,prd,twocolumn,groupedaddress,nofootinbib,floatfix,superscriptaddress]{revtex4-1}
\usepackage{graphicx,bm,color}
\usepackage{aas_macro}
\usepackage{multirow}
\usepackage{verbatim}
\usepackage{ulem}
\usepackage{mathtools}
\usepackage{float}
\usepackage{colortbl}
\usepackage[dvipsnames]{xcolor}
\usepackage{braket}

\usepackage[pdftex,colorlinks,pdfpagelabels]{hyperref}
\hypersetup{
   bookmarksnumbered,
   citecolor={blue},
   linkcolor={blue},
   urlcolor={blue},
   filecolor={blue}
}

\definecolor{pink1}{RGB}{226, 24, 166}

\newcommand{\lsim}{\mathrel{\hbox{\rlap{\lower.75ex \hbox{$\sim$}} \kern-.3em \raise.4ex \hbox{$<$}}}}
\newcommand{\gsim}{\mathrel{\hbox{\rlap{\lower.75ex \hbox{$\sim$}} \kern-.3em \raise.4ex \hbox{$>$}}}}

\begin{document}

\title{Modeling the Optical Cherenkov Signals by Cosmic Ray Extensive Air Showers Directly Observed from Sub-Orbital and Orbital Altitudes}
\author{A.L. Cummings}
\affiliation{Gran Sasso Science Institute (GSSI), L'Aquila 67100, Italy}
\affiliation{INFN-Laboratori Nazionali del Gran Sasso, Assergi (AQ) 67010, Italy}

\author{R. Aloisio}
\affiliation{Gran Sasso Science Institute (GSSI), L'Aquila 67100, Italy}
\affiliation{INFN-Laboratori Nazionali del Gran Sasso, Assergi (AQ) 67010, Italy}

\author{J.Eser}
\affiliation{Department of Astronomy $\&$ Astrophysics, University of Chicago, Chicago, IL 60637, USA}

\author{J.F. Krizmanic}
\affiliation{CRESST/NASA Goddard Space Flight Center, Greenbelt, MD 20771, USA}
\affiliation{University of Maryland, Baltimore County, Baltimore, MD 21250, USA}

\date{\today}

\begin{abstract}
Future experiments based on the observation of Earth's atmosphere from sub-orbital and orbital altitudes plan to include optical Cherenkov cameras to observe extensive air showers produced by high-energy cosmic radiation via its interaction with both the Earth and its atmosphere. As discussed elsewhere \cite{Cummings:2020ycz,Reno:2019jtr}, particularly relevant is the case of upward-moving showers initiated by astrophysical neutrinos skimming and interacting in the Earth. The Cherenkov cameras, by looking above Earth's limb, can also detect cosmic rays with energies starting from less than a PeV up to the highest energies (tens of EeV). Using a customized computation scheme to determine the expected optical Cherenkov signal from these high-energy cosmic rays, we estimate the sensitivity and event rate for balloon-borne and satellite-based instruments, focusing our analysis on the Extreme Universe Space Observatory aboard a Super Pressure Balloon 2 (EUSO-SPB2) and the Probe of Extreme Multi-Messenger Astrophysics (POEMMA) experiments. We find the expected event rates to be larger than hundreds of events per hour of experimental live time, enabling a promising overall test of the Cherenkov detection technique from sub-orbital and orbital altitudes as well as a guaranteed signal that can be used for understanding the response of the instruments. 
\end{abstract}

\maketitle

\section{Introduction}

The observation of the astrophysical flux of high-energy radiation (cosmic rays, gamma rays and neutrinos) is becoming of vital importance in the upcoming era of multi-messenger astronomy (MMA). With the continued detection of gravitational waves \cite{Aloiso:2018hbl}, the ability to be able to measure these high-energy particles in different detection channels reliably and in a timely manner is critical to achieving MMA goals. Among the new experimental techniques being explored to achieve such goals is the focus of this work: the observation of the Earth atmosphere as a vast cosmic ray detector from sub-orbital or orbital altitudes. This technique guarantees huge target masses with an unprecedented increase of experimental exposure. Specifically, we model the detection of the beamed optical Cherenkov light from cosmic-ray induced extensive air showers (EAS) viewed above the Earth's limb.

As in \cite{Cummings:2020ycz,Reno:2019jtr}, we will focus our discussion on the detection capabilities of the pathfinder Extreme Universe Space Observatory aboard a Super Pressure Balloon 2 (EUSO-SPB2) experiment \cite{Wiencke:2019vke}, currently under construction with a targeted launch date in 2023, and the future Probe Of Extreme Multi-Messenger Astrophysics (POEMMA) experiment \cite{Olinto:2020oky,Olinto:2019mjh,Olinto:2017xbi}, both designed to make use of fast imaging (10-20~ns integration time), Schmidt optical Cherenkov telescopes to measure Cherenkov signals from EAS. While EUSO-SPB2 has a dedicated Cherenkov telescope with two 0.35~$\mathrm{m}^{2}$ bifocal mirrors and a 512 pixel Silicon Photo-Multiplier (SiPM) camera, POEMMA is designed with a hybrid focal surface where the Cherenkov telescope portion consists of 15,360 SiPM pixels with an overall 2.5~$\mathrm{m}^{2}$ optical collection aperture. The field of views for the respective Cherenkov cameras are $12.8^{\circ}\times 6^{\circ}$ and $30^{\circ}\times 9^{\circ}$ for EUSO-SPB2 and POEMMA. Both detectors observe the Earth from high altitudes and use the atmosphere as the detector sensitive volume, with EUSO-SPB2 observing from suborbital altitudes, suspended from a super pressure balloon (33~km) and POEMMA observing from orbital altitudes as dual free-flying satellites (525~km).

In their designs, the Cherenkov cameras of both  EUSO-SPB2 and POEMMA target near Earth's limb and below to capture the bright, beamed Cherenkov emission from Earth-skimming neutrino events. As discussed in \cite{Cummings:2020ycz,Reno:2019jtr}, these signals occur when a neutrino passes through the Earth and interacts close enough to the Earth's surface to emerge (bottom-up) into the atmosphere as a muon or $\tau$-lepton. These particles can then decay or interact, generating an upward-moving EAS. For neutrino energies larger than a few PeV, the high-energy charged particles (electron-positron dominant) forming the EAS develop a narrow (opening angle $<1.5^{\circ}$), forward-beamed Cherenkov pulse which can be detected by high-altitude experiments, offering a complimentary methodology of observing high-energy neutrinos to those experiments which measure via radio emission or via optical Cherenkov emission in situ (such as IceCube) \cite{Barwick:2005hn,2009APh....32...10A,Williams:2020mvu,Allison:2011wk,Allison:2015eky,Barwick:2014rca,Barwick:2014pca,Aartsen:2019swn,Adrian-Martinez:2016fdl}. 

As in the case of the Earth-skimming neutrinos, cosmic rays with above-the-limb trajectories and energies larger than a few PeV can interact with Earth's atmosphere and produce an EAS with a resulting optical Cherenkov signal strong enough to be experimentally detectable. Both the EUSO-SPB2 and POEMMA missions are designed such that the onboard Cherenkov camera can provide additional coverage above the Earth's limb, allowing for detection of these cosmic ray events. As we will discuss in sections \ref{sec:traj_char} and \ref{sec:cher_sign}, due to the geometries of the ``above-the-limb" trajectories, much of the development of the particle cascade occurs at high altitude in rarified atmosphere. Because of this, the generation of optical Cherenkov emission is limited, but so too is its atmospheric attenuation during its propagation. In this way, a detailed calculation of the Cherenkov signal strength and geometry is required to determine the overall instrumental sensitivity to such events. 


The interest in studying in closer detail the Cherenkov signal produced by cosmic rays with above-the-limb trajectories is twofold. Firstly, high energy ($\gtrsim$~PeV) cosmic ray events are characterized by higher fluxes (at the level of 4 orders of magnitude) with respect to those of astrophysical neutrinos in the same energy range  \cite{Aloisio:2017ooo}. Secondly, due to the atmospheric refraction of the optical Cherenkov emission, an above-the-limb signal could be reconstructed as a shower originating from below the Earth limb, thereby mimicking a neutrino induced event \cite{Chu:83}. Such a process presents an additional background regarding the observation of the (below-the-limb) neutrino events. An estimate of the rate of the refracted cosmic ray background as a function of detector viewing angle can be useful in developing angular cuts for both EUSO-SPB2 and POEMMA to improve the confidence in measuring an actual Earth-skimming neutrino-induced EAS.

Secondly, as we will discuss in section \ref{sec:cher_sign}, the Cherenkov signal produced by cosmic ray events with above-the-limb trajectories has nearly identical properties to what is expected from below the limb neutrino events, i.e. similar wavelength spectra of arriving photons, as well as similar spatial profiles and time distributions. For this reason and given their expected high rate, above-the-limb cosmic ray events can provide a consistent benchmark to test the different components of a Cherenkov telescope (i.e. optics, electronics and triggers) directly in flight. Thus, a significant statistical sample of these cosmic ray events observed by the optical Cherenkov cameras of sub-orbital and orbital instruments allow for an in-situ determination of the instrumental response to actual EAS Cherenkov signals while also studying their variability.

Moreover, unlike the neutrino events, the optical Cherenkov signals from cosmic rays are directly produced by the primary (cosmic) particle, while for neutrino events, one should take into account the effect of energy losses in the Earth and the decay length/energy spectra of the resulting charged leptons \cite{Cummings:2020ycz,Reno:2019jtr}. Therefore, the reconstruction of the energy of the primary particle in the case of above-the-limb events will be more straightforward and can be used to test the overall Cherenkov detection capabilities of sub-orbital and orbital instruments by providing a measurement of the all-particle cosmic ray spectrum above $\sim$PeV energies.

The observation of cosmic ray events coming from above the limb is not without precedent, having been observed by the Antarctic Impulsive Transient Antenna (ANITA) experiment. ANITA is a balloon-borne instrument designed to detect the high frequency (200-1200~MHz) radio emission produced by neutrinos in ice through the Askaryan effect \cite{Askaryan:1962aa,Askaryan:1989hk} or in the atmosphere through the EAS interaction with the geomagnetic field (with minimal contribution also from the Askaryan radiation). From the second point, ANITA is also sensitive to cosmic rays with a threshold energy for detection around a few EeV. During its four flights, ANITA has detected several UHECR events with reconstructed directions originating from above Earth's limb \cite{Hoover:2010qt,Schoorlemmer:2015afa} in addition to placing stringent limits on the flux of neutrinos with $E_{\nu}>10^{18}$~eV \cite{Gorham:2019guw} 

The analysis presented in this paper is intended as an extension of the computation scheme already discussed in \cite{Cummings:2020ycz}, which determines the optical Cherenkov signal for upward-moving EAS initiated by neutrino interactions in the Earth. Here we extend the computation to include also the above-the-limb events from the high-energy (i.e. starting from energies around a few PeV up to energies above 10~EeV) cosmic ray flux. Assuming a pure proton cosmic ray composition as the reference case, we characterize the properties of the expected Cherenkov signal, and determine the detection rates expected for the EUSO-SPB2 and POEMMA experiments. The paper is organized as follows: in section \ref{sec:traj_char}, we detail the trajectories of the above-the-limb cosmic ray events; in section \ref{sec:cher_sign}, we discuss the Cherenkov emission for the case of proton cosmic rays with energies of PeV and above, and assess the effects of the geomagnetic field on the Cherenkov signal development in high-altitude, rarified atmosphere; in section \ref{sec:aper_sens}, we compute the expected event rate in the case of EUSO-SPB2 and POEMMA experiments; and finally, in section \ref{sec:conc}, we detail our conclusions. 

\section{Above the limb trajectories}
\label{sec:traj_char}

A high-energy cosmic ray impinging the Earth's atmosphere with a trajectory that is directed at an instrument and is viewed above the limb gives rise to an EAS that can spend much of its development at high altitudes, where the atmosphere is rarified. For this reason, it is important to characterize the above-the-limb trajectories to understand the EAS development, Cherenkov emission, and atmospheric attenuation.

\begin{figure}[t!]
\includegraphics[width=\linewidth]{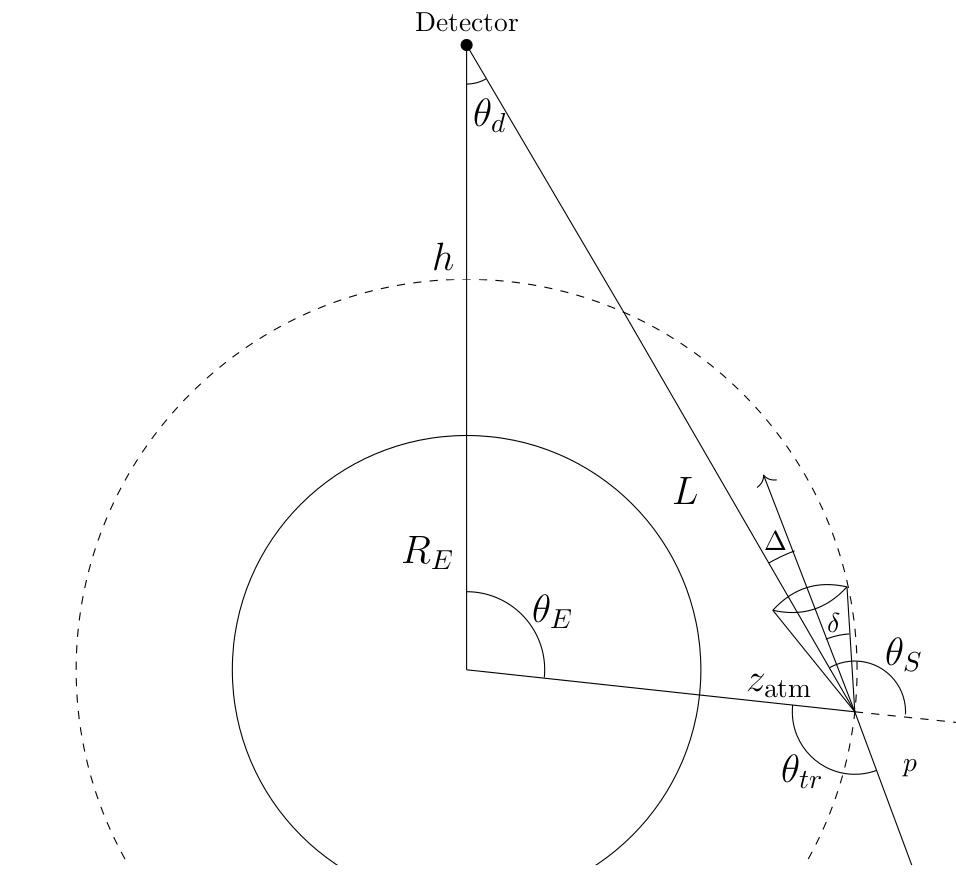}
\caption{Geometry of measuring the Cherenkov signal from cosmic rays arriving from above the Earth horizon in the case of a space based instrument.}
\label{fig:Geometry}
\end{figure}

The typical geometry of an above-the-limb trajectory is sketched in Figure \ref{fig:Geometry} for a satellite-based detector. The distances $h$, $L$, $z_{\mathrm{atm}}$ and $R_{E}$ are the altitude of the detector, the path length traveled by the shower front, the height of the atmosphere at the cosmic ray impinging point, and the Earth radius. As in \cite{Cummings:2020ycz}, we use the 1976 US standard atmosphere to describe the atmospheric density as a function of altitude, taking the top of the atmosphere to be at $z_{atm} \simeq 113$~km \cite{USatmo:1976aaa}. 

The three angles $\theta_{S}$, the angle of the detector's optical axis (with respect to the local zenith), $\theta_{d}$ the detector's viewing angle with respect to nadir, and $\theta_{E}$, the Earth viewing angle with respect to the center of the Earth, are related as $\theta_d=\theta_s-\theta_E$. The angle $\Delta$ is the angular difference between the particle trajectory and the detector's line of sight and $\delta$ is the angle off of the shower axis within which the Cherenkov emission can be experimentally detected. The EAS trajectory is defined by the angle $\theta_{tr}$ with respect to the local zenith at the impinging point. 

Referring to Figure \ref{fig:Geometry}, it easily follows that the observable trajectories above the Earth's limb are bracketed inside the detector viewing angle range:

\begin{equation}
\begin{split}
\mathrm{sin}^{-1}\left(\frac{R_{E}}{R_{E}+h}\right) < \theta_{d} <
\begin{cases}
\frac{\pi}{2} &h<z_{\mathrm{atm}}\\
\mathrm{sin}^{-1}\Big(\frac{R_{E}+z_{\mathrm{atm}}}{R_{E}+h}\Big) \, &h > z_{\mathrm{atm}}
\end{cases}
\end{split}
\end{equation}
where the maximum viewing angle differs in the case of an instrument placed inside the atmosphere or outside of it.

The $\frac{\pi}{2}$ limit in the suborbital case is somewhat arbitrary, disallowing for events which have downwards trajectories. We later demonstrate that this is a sufficient limit, as the limited thickness of the atmosphere at balloon altitudes disallows for significant optical Cherenkov emission. Following this, in the case of EUSO-SPB2, the above-the-limb trajectories can be geometrically bracketed inside the viewing angle range 
$84.2^{\circ}<\theta_{d}<90^{\circ}$; while in the case of POEMMA, the corresponding viewing angle range shrinks to $67.5^{\circ}<\theta_{d}<70^{\circ}$. The angular ranges given here are purely geometrical restrictions and do not consider the total grammage of the atmosphere along these trajectories in which a cosmic ray can interact. Taking this point into account will further reduce the angular range (see below).

\begin{figure}[t!]
\includegraphics[width=\linewidth]{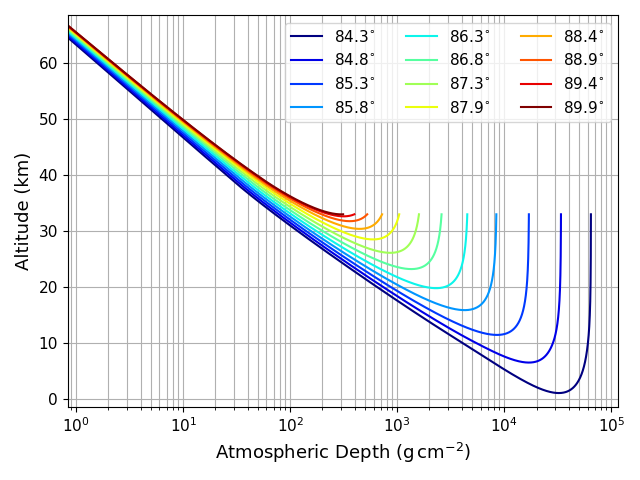}
\includegraphics[width=\linewidth]{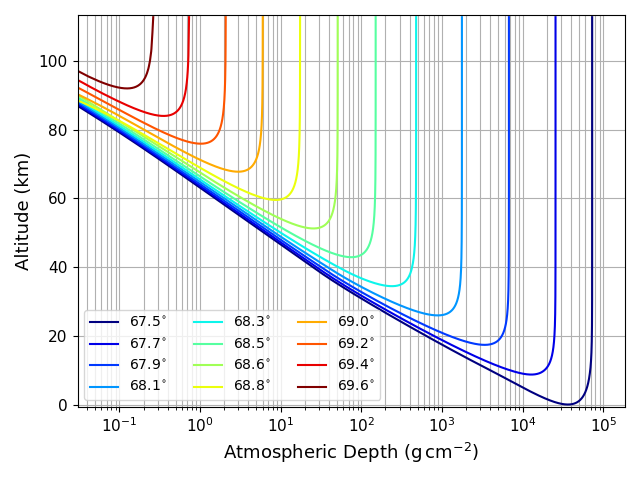}
\caption{Cumulative slant depth as a function of altitude and detector viewing angle from nadir, as measured from 33~km altitude [upper panel] and 525~km altitude [lower panel]. The Earth limb appears at $\theta_{d} = 84.2^{\circ}$ and $\theta_{d} = 67.5^{\circ}$ for 33~km and 525~km observation altitudes, respectively. Calculations assume the 1976 US standard atmosphere \cite{USatmo:1976aaa}.}
\label{fig:grammage}
\end{figure}

The cumulative slant depth as a function of path length traveled by a particle through the atmosphere can be found by integrating the atmospheric density along the particle trajectory for a given detector viewing angle. Assuming the 1976 US standard atmosphere \cite{USatmo:1976aaa}, the slant depth profiles for the observation altitudes of EUSO-SPB2 (33~km) and POEMMA (525~km) are plotted in Figure \ref{fig:grammage} across the labeled viewing angles. 

Cosmic ray air showers complete their full development over a distance of roughly $ 1000 \, \mathrm{g} \, \mathrm{cm}^{-2}$, with the shower maximum $X_{\mathrm{max}}$ (the slant depth where maximum shower development occurs, and thus a good estimate of the overall shower properties) occurring from $\sim 500 \, \mathrm{g} \, \mathrm{cm}^{-2}$ to $\sim 800 \, \mathrm{g} \, \mathrm{cm}^{-2}$, depending on the primary energy and mass composition of the cosmic ray \cite{Aloisio:2017ooo}. Analyzing the trajectories presented in Figure \ref{fig:grammage}, we can begin to quantify the regions of characteristic shower development. By taking a representative shower $X_{\mathrm{max}}= 700 \, \mathrm{g} \, \mathrm{cm}^{-2}$ corresponding to the case of a 100~PeV proton, we observe that the altitude of maximum shower development has a minimum of $\sim20$~km for both balloon and satellite trajectories, which increases with increasing viewing angle, indicating the need to carefully account for shower development at high altitudes.



The first interaction point (the point of EAS initiation) is distributed exponentially with a mean interaction length of $\lambda$. For a proton primary, $\lambda$ decreases from roughly $70~ \mathrm{g} \, \mathrm{cm}^{-2}$ at 1~PeV energy to $40~ \mathrm{g} \, \mathrm{cm}^{-2}$ at 10~EeV. This implies that for large viewing angles, where the atmosphere is thin, large variations in the optical Cherenkov signal are expected due to the shower-to-shower fluctuations and it will be necessary to consider also this effect.

\section{Optical Cherenkov Signal}
\label{sec:cher_sign}
In this section, we discuss the main characteristics of the optical Cherenkov emission produced by electron-positron pairs (hereafter electrons) in EAS initiated by cosmic rays with above-the-limb trajectories. The Cherenkov emission produced from secondary muons in the EAS is not considered in this work but can become especially important when considering mass composition estimates, see \cite{Neronov_2016,krolik2019cherenkov}. The contribution of the secondary muons to the total Cherenkov signal can potentially strengthen the probability of observing the EAS, particularly where the atmospheric slant depth of the cosmic ray trajectory is large. by considering only the emission generated by electrons and positrons, we provide a conservative estimate for the strength of the signal.

The computation scheme that we use to generate these signals was originally developed in \cite{Cummings:2020ycz} to determine the optical Cherenkov signal produced by upward-moving EAS observed below the Earth's limb sourced from neutrino interactions in the Earth, taking into account the evolution of the electron energy, angular, and lateral distributions. For simplicity, we will not review all aspects of this computation scheme, but discuss the key points specifically relevant (or different) for Cherenkov signals produced by above-the-limb cosmic ray induced EAS. Specifically, we will concentrate on how EAS development in high altitude environments is handled differently and what the resulting effects are on the properties of the simulated Cherenkov light generation and detection.

 
A required input for the optical Cherenkov simulation is the particle trajectory, that is, the description of cumulative slant depth as a function of propagated distance, shown by the curves presented in Figure \ref{fig:grammage}. To this purpose we have slightly modified the computation scheme of \cite{Cummings:2020ycz}, determining the EAS properties along the path-length $L$ instead of along the altitude $z$, as it follows from the fact that in the case of above-the-limb trajectories, the altitude no longer uniquely tags the EAS properties.

As we have shown in section \ref{sec:traj_char}, for these above-the-limb trajectories, much of the Cherenkov generation in an EAS occurs at altitudes $>20$~km, where we have previously detailed in \cite{Cummings:2020ycz} that certain effects become relevant (particularly the lateral distribution of the generating electrons, which scales with the Moliere radius, being $r_{m} \sim 1$~km for $z=20$~km) and therefore require careful tracking and handling.
 
The index of refraction $n(z)$ in the Earth atmosphere decreases exponentially with increasing altitude \cite{Bernlohr:2008mpk,Weast:1986uoy,Ciddor:1996aaa}. As such, there is a corresponding decrease in the local Cherenkov angle $\theta_{ch}=\cos^{-1}(1/ \beta n(z))$ from $\sim 1.4^{\circ}$ near sea level to $<0.3^{\circ}$ at the characteristic altitudes of above-the-limb EAS development. Therefore, we can expect that the typical angular scales of the Cherenkov emission from the above-the-limb events are smaller than those of the neutrino induced EAS.

In addition, the decreased index of refraction reflects in a high Cherenkov energy threshold, which fixes the energy above which electrons can radiate through the Cherenkov effect (i.e. when $\beta>n(z)$) namely:
 
\begin{equation}
E>E_{thr} = \frac{m}{\sqrt{1-\frac{1}{n(z)^{2}}}}
\label{eq:en_threshold}
\end{equation}

\begin{figure}[t!]
\includegraphics[width=\linewidth]{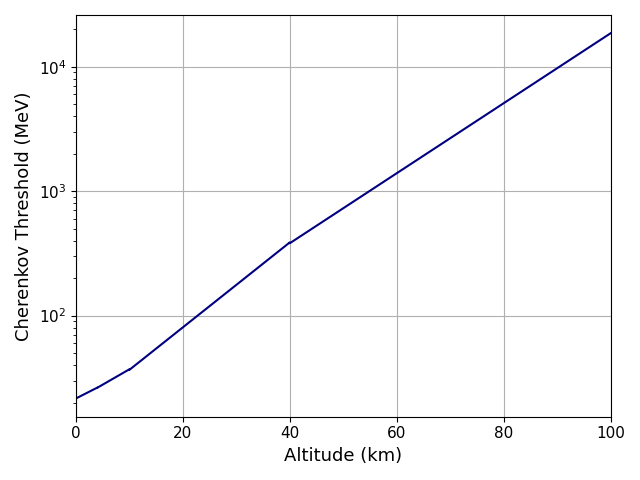}
\caption{Cherenkov threshold of electrons as a function of altitude for $\lambda$=450~nm. Dispersion effects are negligible in our Cherenkov wavelength range 200~nm to 1000~nm.}
\label{fig:threshold}
\end{figure}

\begin{figure}[t!]
\includegraphics[width=\linewidth]{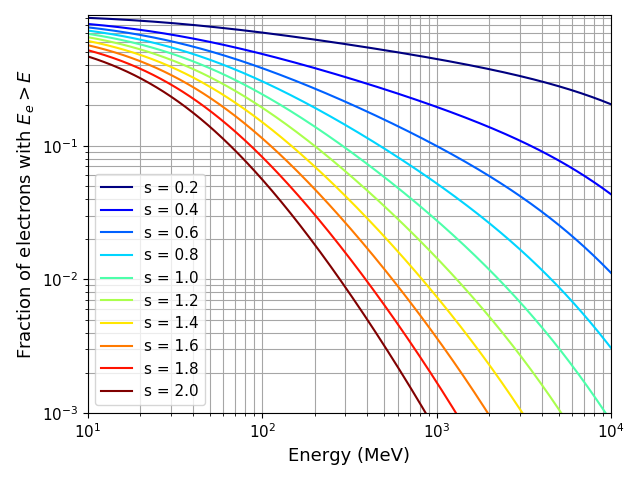}
\caption{Fraction of $e^{\pm}$ above energy $E$ for various shower ages, as parameterized in \cite{Hillas:1982wz}}
\label{fig:hillas_energy_threshold}
\end{figure}

In Figure \ref{fig:threshold}, we plot the Cherenkov threshold energy of electrons (expressed in MeV) as a function of altitude, assuming a central Cherenkov emission wavelength of 450~nm. In Figure \ref{fig:hillas_energy_threshold}, we plot the fraction of electrons above energy $E$ (expressed in MeV) in a given EAS as a function of the shower age $s$ ($s = 3/[1+2(X_{\mathrm{max}}/X)]$), as parameterized by Hillas \cite{Hillas:1982vn}.
 
At characteristic altitudes for above-the-limb trajectories, the relevant Cherenkov thresholds range from $\sim100$~MeV up to $\sim1$~GeV, as shown in Figure \ref{fig:threshold}. Notwithstanding these high-energy thresholds, as follows from Figure \ref{fig:hillas_energy_threshold}, there is still a significant fraction of electrons within the EAS able to emit Cherenkov light, especially in the case of early shower ages. For instance, taking the case of a shower maximum occurring at 30~km altitude, where the Cherenkov threshold is around $200$~MeV, $15\%$ of the electrons in the EAS are still energetic enough to produce Cherenkov emission, with even higher fractions of electrons contributing to the emission earlier in the shower. As the generation of Cherenkov photons occurs on electrons with characteristically higher energies for above-the-limb EAS with respect to EAS induced by Earth-skimming neutrinos, there will be a further reduction in the angular scales of the resulting emission as higher energy electrons are deflected less from the shower axis.

If compared to the case of upward-moving, Earth-interacting, neutrino generated EAS, the reduced number of emitting electrons in an above-the-limb cosmic ray induced EAS produces marginally weaker Cherenkov emission. On the other hand, the high altitudes at which these EAS develop guarantees a lower atmospheric attenuation of the generated Cherenkov photons due to the lower Rayleigh and aerosol path lengths. These two competitive effects primarily determine the intensity of the signal observed at sub-orbital and orbital altitudes, while the angular and lateral properties of the electrons shape the spatial distribution of the arriving photons.

In order to quantify the impact of the atmospheric attenuation, following the approach of \cite{Cummings:2020ycz}, we take into account the effects of the Cherenkov photon propagation through the atmosphere for the above-the-limb trajectories, using the atmospheric extinction models from \cite{Elterman:1968,Blitzstein:1970}. In this model, the atmospheric extinction of visible light, given in the wavelength range 270~nm to 4000~nm, is primarily due to Rayleigh, aerosol and ozone scattering, which vary as a function of altitude, and are given within the range 0~km to 50~km. For altitudes above 50~km, we assume only the effects of the Rayleigh scattering, as the concentration of aerosols and ozone are small and the effects are subsequently minimized (see tables in \cite{Elterman:1968}). We will later show that the typical above-the-limb cosmic ray EAS which are observed by balloon-borne and space-based instruments have trajectories which do not allow for strong development above 50~km altitudes, further validating this approximation.

\begin{figure}[t!]
	\includegraphics[width=\linewidth]{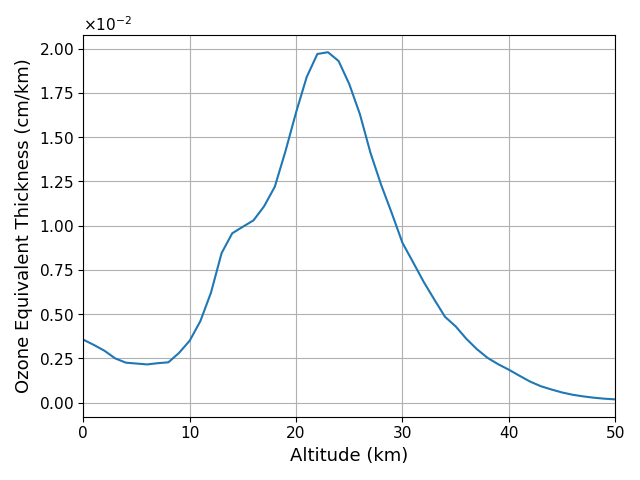}
	\includegraphics[width=\linewidth]{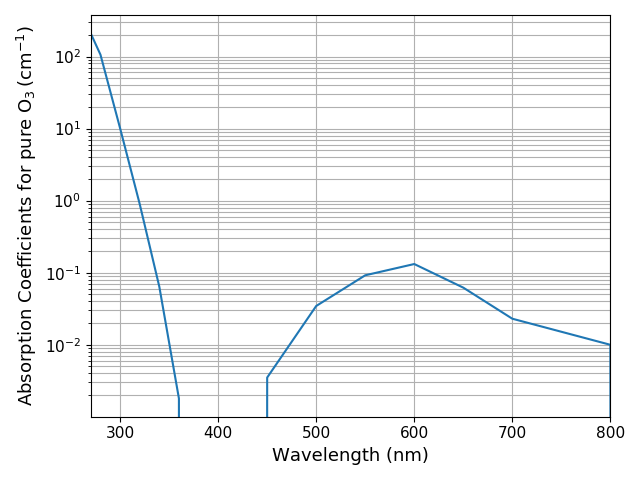}
	\caption{[Upper panel] Equivalent Ozone thickness as a function of altitude. [Lower panel] Ozone absorption coefficient as a function of wavelength. Both data-sets are taken from \cite{Elterman:1968}. The strong decrease in the absorption coefficient near $\lambda = 400$~nm is consistent with more recent measurements of the ozone-light cross section \cite{AMT}.}
	\label{fig:Ozone}
\end{figure}

Particularly relevant for the above-the-limb trajectories is the altitude range from 15~km to 35~km, where a significant resurgence in the ozone concentration is observed. Ozone strongly attenuates light with wavelength below 300~nm, with a minimum attenuation around 400~nm and a subsequent increase with a local maximum around 600~nm. The ozone effective thickness as a function of altitude $D(z)$ from \cite{Elterman:1968} is shown in the upper panel of Figure \ref{fig:Ozone} while the lower panel shows the Ozone absorption coefficients as a function of wavelength $A(\lambda)$ (proportional to the cross section per molecule of the interaction between photons and ozone molecules). The Ozone attenuation coefficient is calculated as $\alpha(z,\lambda) = D(z)A(\lambda)$.

Showers with above-the-limb trajectories develop significantly at the altitudes where ozone attenuation is prevalent, and we therefore expect to observe the imprint of the light-ozone cross section in the wavelength behavior of the Cherenkov emission spectra observed by sub-orbital and orbital altitudes.

\begin{figure}[t!]
\includegraphics[width=\linewidth]{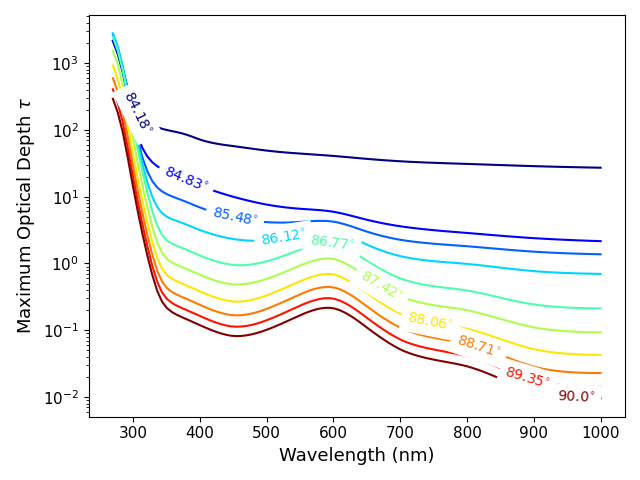}
\includegraphics[width=\linewidth]{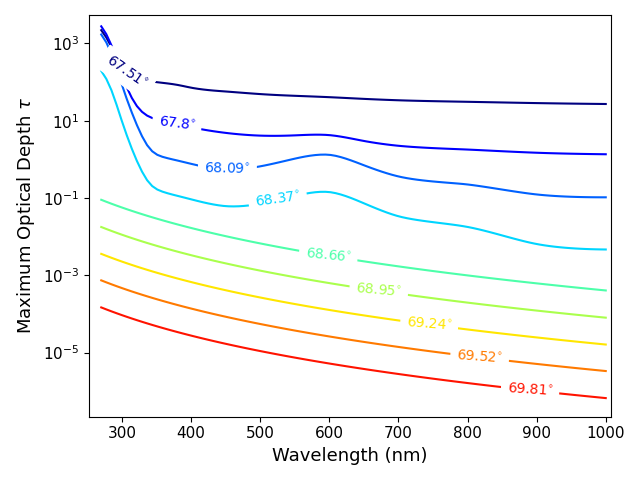}
\caption{Maximum optical depth as a function of wavelength for different shower trajectories for 33~km altitude observation [upper panel] and 525~km altitude observation [lower panel].}
\label{fig:Optical_Depth}
\end{figure}

Following the approach of \cite{Cummings:2020ycz}, in Figure \ref{fig:Optical_Depth}, we plot the maximum optical depth $\left(\tau = \int_{L=0}^{L_{det}}\alpha(z)dL\right)$ of the total atmospheric extinction as a function of wavelength and for different detector viewing angles assuming, in the upper panel, the observation altitude of EUSO-SPB2 (33~km) and, in the lower panel, of POEMMA (525~km). In both cases, the atmospheric extinction for trajectories that point toward the Earth's limb are dominated by low altitude effects, that is, the Rayleigh and aerosol scattering. For this reason, at these angles, we do not strongly observe the features of the ozone absorption.  

However, in the case of a balloon-borne instrument at 33~km altitude, e.g. EUSO-SPB2, for high viewing angles (up to 90$^\circ$ from nadir), the trajectories through the atmosphere correspond to altitudes $>10$~km, as follows from the upper panel of Figure \ref{fig:grammage}, and the effect of the ozone layer becomes more relevant. This is observed in the upper panel of Figure \ref{fig:Optical_Depth} where the optical depth corresponding to high viewing angles shows a local maximum at 600~nm, a minimum around 400~nm and a strong increase towards lower wavelengths, as follows from the behavior of the light-ozone cross section.  

In the case of a satellite-based instrument at 525~km altitude, e.g. POEMMA, the effect of the ozone layer is relevant only within a small angular range. For angles very close to the limb, the Rayleigh and aerosol scattering dominate, while trajectories above the limb by ~1$^\circ$ or more do not experience altitudes below 50~km, as shown in the lower panel of Figure \ref{fig:grammage}, and thus experience only Rayleigh scattering under the assumptions of our atmospheric modeling. This is observed in the optical depth behavior in the lower panel of Figure \ref{fig:Optical_Depth}, where the familiar features of the ozone scattering are relevant only for a small range of trajectories.

\subsection{Geomagnetic Field Effects}
\label{sec:geofield}
As previously stated, the dominant contribution to the generation of Cherenkov photons during EAS development is given by the electrons and positrons due to their high abundance and decreased Cherenkov threshold with respect to the other charged particle species in the shower \cite{Gaisser:2016uoy}. In principle, Earth's geomagnetic field can potentially affect the shower development both through synchrotron emission, which reduces the energy of the electron-positron pairs, and through the angular deflection of electrons and positrons from one another by the Lorentz force, which also affects the angular distribution and the lateral spatial size of the shower front. The length scales of these processes are dependent on the electron energy and the strength of the magnetic field, which, in the Earth atmosphere, is effectively constant below 100~km altitudes, where EAS development occurs (the strength of the geomagnetic field scales as $1/R^{3}$, with $R$ the distance from the center of the Earth), changing magnitude and orientation only with geographic position. Typical values of the geomagnetic field strength range from roughly 25~$\mu$T at the equator up to 65$~\mu$T at the poles \cite{GeoField}.

The synchrotron energy losses suffered by electrons and positrons become important only for energies larger than $E>100$~GeV \cite{Hillas:1982vn,Homola:2014sra,Bartoli:2014sza}. As shown in Figure \ref{fig:hillas_energy_threshold}, these energies correspond to an extremely limited number of electrons in the EAS. Therefore, in the forthcoming discussion, we neglect the effect of synchrotron energy losses.

The only relevant effect of the geomagnetic field on the EAS development with respect to the optical Cherenkov emission concerns the deflection of electrons and positrons. The gyration radius of an electron at the Cherenkov threshold in Earth's magnetic field scales roughly as $r_{g} \propto \rho(z)^{-1/2}$ following the discussion of the Cherenkov energy threshold in equation \ref{eq:en_threshold}. The radiation length (the grammage required for a typical electron to lose a fraction $e^{-1}$ of its energy through ionization and Bremsstrahlung emission) measures $X_{r}=37 \, \mathrm{g} \, \mathrm{cm}^{-2}$ in air and has a corresponding linear distance that scales as $L_{r} \propto \rho(z)^{-1}$. These length scales become comparable above altitudes of 25~km, allowing for effective separation of electrons and positrons. 


The behavior of above-the-limb EAS interacting with the geomagnetic field differs from the case of upward going neutrino induced EAS. As discussed in \cite{Cummings:2020ycz,Olinto:2020oky}, it is unlikely for a neutrino sourced EAS to develop significantly at high altitude. High energy charged leptons sourced by neutrino interactions in the Earth are preferentially produced very close to the Earth's limb, and the extended decay length is balanced by the nearly horizontal shower trajectory. Therefore, ignoring the effect of the geomagnetic field was a reasonable, initial approximation for the analysis detailed in \cite{Cummings:2020ycz}. 

For cosmic ray events arriving from above the limb, the majority of the shower develops above altitudes of 20~km (see Figure \ref{fig:grammage}), where the atmosphere is thin and the interaction distance is large. Thus, the deflection of electrons and positrons in the shower may not be trivial and should be considered. 

To quantify the effect of the geomagnetic field, in what follows, we use a simple phenomenological approach that quantifies the maximal effect of the electron-positron pair deflections on the optical Cherenkov emission. In this approach, we ignore the cumulative bending of the generating electrons and positrons in the geomagnetic field, taking instead the electron angular distributions as given in \cite{Hillas:1982vn} when calculating the differential deflection of the resultant Cherenkov emission. This first order approximation is sufficient, as the majority of the electrons which generate Cherenkov emission have energies larger than the Cherenkov threshold, thereby reducing the significance of the magnetic bending with respect to electron energy loss processes (see Figure \ref{fig:hillas_energy_threshold}). For a detailed discussion regarding the effects of the geomagnetic field on optical Cherenkov emission generated by downwards going EAS, see also \cite{Homola:2014sra}. 

The trajectories of the electron-positron pairs in the EAS are deflected by the geomagnetic field in the plane perpendicular to the field. On this plane, the particle trajectory is circular with a radius $r_g$ (Larmor radius) given by:

\begin{equation}
r_g = \frac{\gamma m \|\vec{v}\|}{q\| \hat{v} \times \vec{B} \|}
\label{eq:rg}
\end{equation}

where $v$ is the particle velocity, $q$ is the particle's charge, $B$ is the magnetic field strength, and $\gamma$ and $m$ are the particle's Lorentz factor and mass. 

\begin{figure}[t!]
	\includegraphics[width=\linewidth]{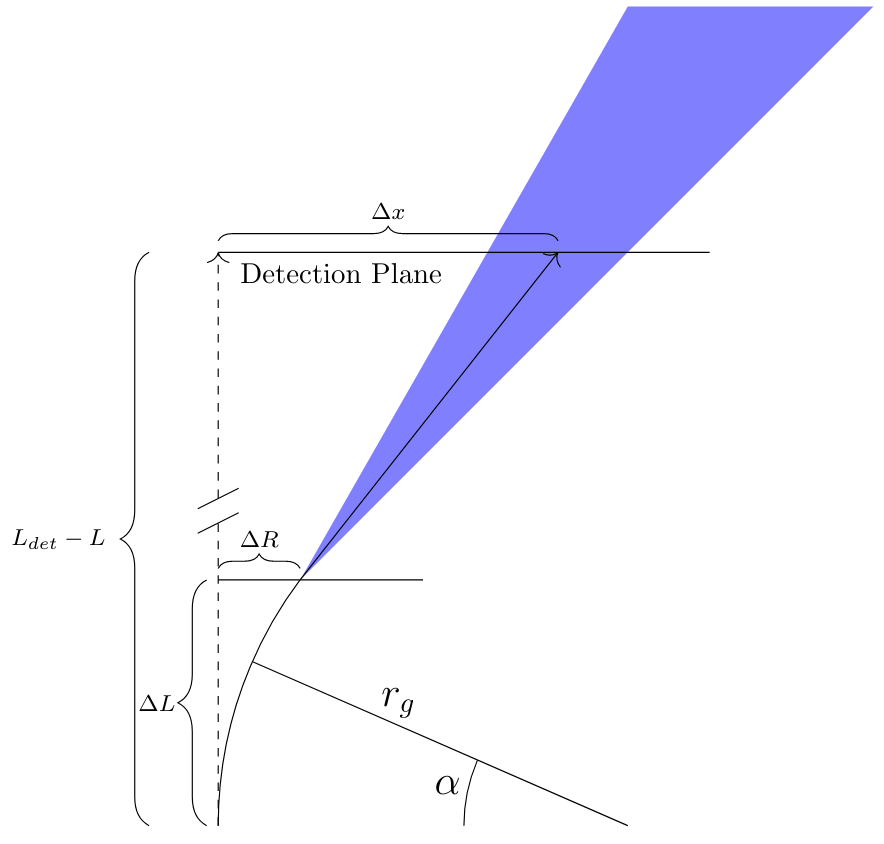}
	\caption{Diagram representing the geomagnetic deflection of electrons and positrons and the effect on the optical Cherenkov emission. The dashed line represents the initial trajectory of the charged particle with no magnetic field.}
	\label{fig:Geo_deflection_diagram}
\end{figure}

The actual distance covered by the electrons in a given bin in slant depth $\Delta X$ ($\Delta L = \Delta X/ \rho(z)$) will be increased by the gyration around $\vec{B}$. In other words, the net effect of the geomagnetic field is to add an offset $\Delta R$ to the electrons along the plane perpendicular to the magnetic field. A diagram of the geomagnetic deflection of electrons is shown in Figure \ref{fig:Geo_deflection_diagram}.

Orienting in the plane perpendicular to the applied magnetic field as outlined in Figure \ref{fig:Geo_deflection_diagram}, the charged particle trajectory can be described via the coordinates:

\begin{equation}
(x, y)=(r_{g}(1-\cos \alpha), r_{g} \sin \alpha)
\end{equation} 

Solving for $\Delta R$ at $y=\Delta L$ yields:

\begin{equation}
\Delta R = r_{g} \left(1-\sqrt{1-\left(\frac{\Delta L}{r_{g}}\right)^{2}} \right)
\label{eq:dltrue}
\end{equation}

 Assuming small angular deviations ($\Delta L \ll r_{g}$), i.e. small enough slant depth bin $\Delta X$, and relativistic particles $\beta\to 1$ one has:

\begin{equation}
\Delta R= \frac{\Delta L^{2}}{2 r_{g}}
\label{eq:dL}
\end{equation}

In our computation scheme, each electron bunch is assigned an energy $E$, a zenith direction $\theta_{e}$ as given by the distributions presented in \cite{Hillas:1982wz,Lafebre:2009en} and an azimuth direction $\phi_{e}$ assigned randomly from 0 to $2\pi$. 

In order to consider the maximal effect of the geomagnetic field, in what follows, we orient the initial magnetic field perpendicular to the EAS propagation direction ($\hat{z}$ axis) along the $\hat{y}$ axis, using a quite high magnetic field strength $B_0=50$ $\mu$T. To calculate the cross term in equation \ref{eq:rg}, we take $\hat{v}_e=(\mathrm{sin}\theta_{e} \mathrm{cos}\phi_{e},\mathrm{sin}\theta_{e} \mathrm{sin}\phi_{e},\mathrm{cos}\theta_{e})$ and $\vec{B}=(0,B_0,0)$, obtaining: $\| \hat{v_{e}}\times \vec{B} \| = B_0 \sqrt{\cos^{2}\theta_{e}+\sin^{2}\theta_{e}\cos^{2}\phi_{e}}$.

After calculating $\Delta R$ using equation \ref{eq:dltrue} for each electron bunch, the electron offset angle within the bin $\Delta L$ which must be projected to the detection plane is calculated as:

\begin{equation}
\theta_{m} = \mathrm{tan}^{-1}(\Delta R/\Delta L)~.
\label{eq;dL}
\end{equation}

The additional offset of the Cherenkov photons on the detection plane due to the deflection of the electrons in the EAS is given by:

\begin{equation}
\Delta x = (L_{det}-L)\mathrm{tan}\theta_{m}
\end{equation}

where $L$ is the distance along the shower axis from the shower starting point to the Cherenkov emission point, and $L_{det}$ is the distance from the shower starting point to the detection plane. This additional distance off axis for each electron bunch is applied randomly in the positive and negative $\hat{x}$ direction to account for the opposite electron and positron charges. In this paper, the geomagnetic field deflections are treated neglecting the EAS charge asymmetry, i.e. assuming an equal number of electrons and positrons. This effect is maximally around $20\%$, but is significantly smaller for the electron energies characteristic of in-air Cherenkov emission ($>20$~MeV) \cite{Gaisser:2016uoy,Lafebre:2009en}.

\begin{figure}[t!]
\includegraphics[width=\linewidth]{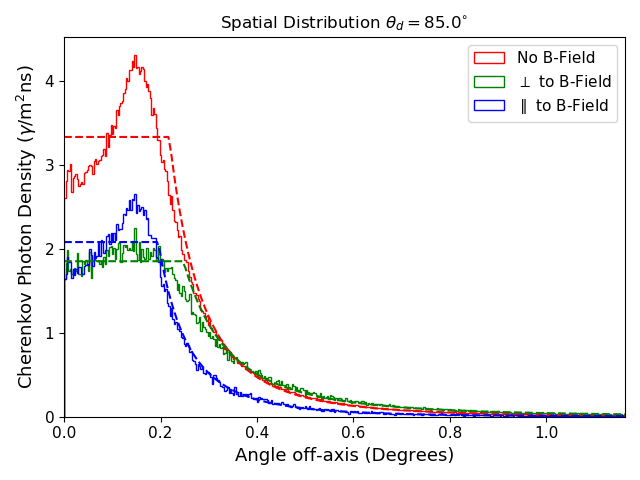}
\caption{Cherenkov spatial distributions of a 100~PeV proton shower as observed from 33~km with $\theta_{d}=85^{\circ}$. The Cherenkov photon distributions are computed in the case where no geomagnetic field is applied (red curve) and where a 50$\mu$T magnetic field is applied perpendicular to the shower propagation direction. The green and blue curves show the components of the spatial distribution measured along the axes perpendicular and parallel to the magnetic field, respectively. The dashed lines show the result of the 5 parameter fit as described by equation \ref{1dprof}.}
\label{fig:Mag_Field_Effects}
\end{figure}

An example of the Cherenkov photon spatial distribution under the effect of the geomagnetic field is shown in Figure \ref{fig:Mag_Field_Effects}. We cut our 3-dimensional distribution along the axes parallel (blue curve) and perpendicular (green curve) to the magnetic field in order to evaluate the maximum and minimum effects of the field. As a comparison, we also show the case where no magnetic field is applied (red curve). From Figure \ref{fig:Mag_Field_Effects}, we observe that the effect of the geomagnetic field is to spread the Cherenkov photons over a wider angular range along the axis perpendicular to the field with respect to the case where no magnetic field is applied. Similarly, as expected, the shape of the profile along the axis parallel to the field is largely the same as that of the unmodified profile, with only the absolute intensity decreasing due to photons within the ring spreading along the perpendicular axis.

\subsection{Time Spread of Arriving Photons}
Cosmic rays interacting in the Earth atmosphere can deposit most of their energy into the ensuing EAS. For this reason, the optical Cherenkov signals from cosmic ray events arriving from above the limb can be extremely bright, and, in principle, visible far off the shower axis, leading to large geometric apertures and high estimated event rates. However, we expect the Cherenkov photons arriving on a given detection plane to have a larger range of arrival times the further off-axis they are observed \cite{Otte:2018uxj,Hillas:1982vn,Homola:2014sra}.

In a general sense, the integration time of a given instrument is typically characterized by the time scale of an expected signal so as to optimize the signal to noise ratio (SNR) with respect to given backgrounds. In the case of the Cherenkov cameras within the framework of the EUSO-SPB2 and POEMMA designs, the integration time is respectively 10~ns and 20~ns, corresponding to the time width of a typical Cherenkov pulse observed close to the shower axis, where the signal is maximized.

By viewing the optical Cherenkov signal further away from the shower axis, the time spread of the arriving photons may be larger than the integration time of the instrument, and the signal is not fully captured, effectively reducing the detection capability. A good description of the photon arrival time is required in order to estimate this reduction and provide a more realistic detector response. While we have a good understanding of the time spread of the optical Cherenkov signal from downward-going showers \cite{Hillas:1982vn,Homola:2014sra}, it is necessary to quantify these features also for upward-moving showers with above-the-limb trajectories.


There are two main propagation time scales involved in this calculation: (i) the propagation time of the electrons in the shower from their origin to their Cherenkov emission point and (ii) the propagation time of photons from the emission point to the detection plane. 

\begin{figure}[t!]
\includegraphics[width=\linewidth]{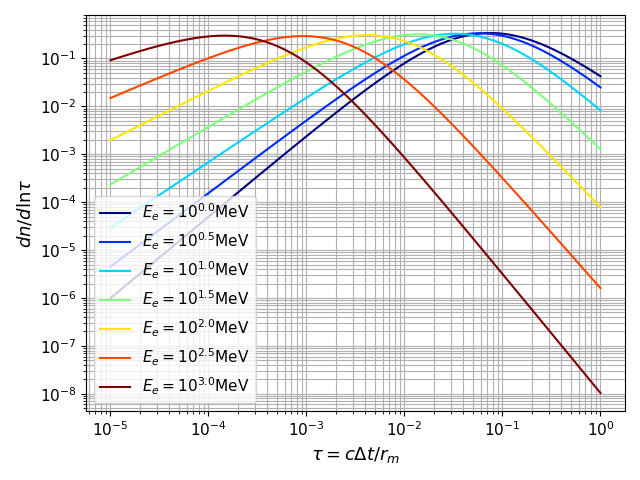}
\includegraphics[width=\linewidth]{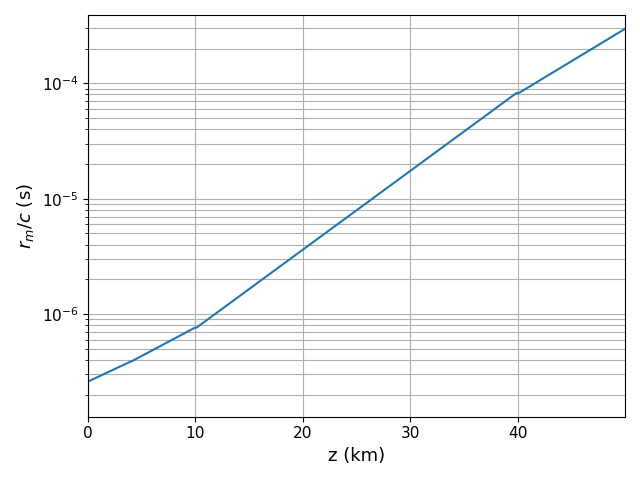}
\caption{[Upper panel] Normalized electron delay time distributions for the scaled variable $\tau=c \Delta t/r_{m}$ for electron energies from 1~MeV to 1~GeV at shower maximum $\tilde{s}=0$ \cite{Lafebre:2009en}. The threshold energy for optical Cherenkov generation is minimally 20~MeV and increases with altitude. [Lower Panel] Moliere radius divided by the speed of light $r_{m}/c$ as a function of altitude using the model of the standard US atmosphere \cite{USatmo:1976aaa} to describe the atmospheric density profile.}
\label{fig:LafebreDist_time}
\end{figure}

To estimate the first time scale, we start by assuming that the shower front moves at the speed of light $c$ so that we can determine the lower bound $ t_0 = L/c$, where $L$ is again the distance from the shower starting point to the emission point, measured along the shower axis. Referring to the computation scheme discussed in \cite{Cummings:2020ycz}, we calculate $t_0$ for every point measured along the charged particle longitudinal profile. We then correct for the fact that the propagating electrons in the EAS reach a plane at a given linear distance along the shower trajectory at different times due to relativistic effects, scattering, and the non-planar shower front. To model these effects, we use the electron time delay distributions (that is, how much the electrons lag an ideal particle traveling at the speed of light) as a function of the electron energy and shower age as provided by Lafebre et al. \cite{Lafebre:2009en}. There is poor universality of the arrival time distribution within the shower over many altitudes and the introduction of a scaled dimensionless variable $\tau$ is helpful:

\begin{equation}
\tau = \frac{c \Delta t}{r_{m}}
\label{eq:lafebre_time}    
\end{equation}
where $\Delta t$ is the delay time and $r_{m}$ is the Moliere radius, which exponentially increases as a function of altitude. The form of the distribution is given as \cite{Lafebre:2009en}:

\begin{equation}
\begin{split}
\frac{dn}{d\mathrm{ln}E d \mathrm{ln} \tau} &= C x^{\zeta_{0}''}(\tau_{1}'+\tau)^{\zeta_{1}''}\\
\tau_{1} &= \mathrm{exp}[-2.71+0.0823\mathrm{ln}E-0.114\mathrm{ln}^{2}E]\\
\zeta_{0}'' &= 1.70+0.160\tilde{s}-0.142 \mathrm{ln}E\\
\zeta_{1}'' &= -3.21
\end{split}
\label{eq:Lafebre_time_dist}
\end{equation}
where $C$ is a normalization constant, $E$ is the electron energy in MeV, and $\tilde{s}$ is a modified shower age $\tilde{s} = (X-X_{\mathrm{max}})/X_{0}$, $X_{0}=36.7 \mathrm{g}/\mathrm{cm}^{2}$ being the radiation length in air. The normalized electron delay time distribution for the dimensionless variable $\tau$ is plotted in the upper panel of Figure \ref{fig:LafebreDist_time} within the electron energy range (1~MeV, 1~GeV) at shower maximum $s=1.0$ with the scale of the process (Moliere radius $r_{m}$ divided by the speed of light $c$) as a function of altitude plotted in the lower panel. As a point of reference, we note that a $\tau$ value of 1 corresponds to a delay time compared to the speed of light $\Delta t$ of 0.26$\mu$s at sea level.

With increasing electron energy, we observe smaller delay times (that is, times which are associated with speeds close to the speed of light), as expected. Additionally, for showers which develop at high altitudes, we observe larger delay times, regardless of electron energy. Using the terminology developed in \cite{Cummings:2020ycz}, for each bunch of photons sampled along the projected ellipse of the Cherenkov ring, we sample a time delay from the above distributions, given an electron energy and shower age. 

The total time for electrons in the EAS to propagate from the shower starting point to the Cherenkov emission point is given as: $t_{e}(L,E,s)=t_{0}(L)+\Delta t(z(L),E,s)$, being $t_0$ the lower bound to the electrons propagation time. 

The second relevant time-scale in calculating the time spreading of the arriving signal is the propagation time of photons through the Earth atmosphere. For each point along the shower, the time needed by photons to travel from the emission point to the detection plane directly along the shower axis can be calculated as:

\begin{equation}
t_{\gamma}(L) = \int_{L}^{L_{det}} \frac{n(z(L))}{c} dL
\label{eq:index_of_refraction_time}
\end{equation}

where $n(z(L))$ is the local refraction index of air. The value of $n$ does not vary strongly across the wavelength range 300~nm to 1000~nm \cite{Bernlohr:2008mpk,Weast:1986uoy,Ciddor:1996aaa}, so we use a central value of $n_{450\mathrm{nm}}(z)$ to further simplify the calculations.

To take into account the effects of off-axis propagation for each point on the Cherenkov ring, we make the approximation $t_{\gamma}(L,\theta_{\gamma}) \approx \frac{1}{\mathrm{cos}\theta_{\gamma}} t_{\gamma}(L)$, where $\theta_{\gamma}$ corresponds to the propagation angle of generated photons through the atmosphere with respect to the shower propagation axis, considering also the lateral spreading of the EAS. We note that $\theta_{\gamma}$ is small for most circumstances, even taking into account the effect of the geomagnetic field.


Our approximation assumes that the index of refraction as a function of $L$ does not change dramatically moving away from the propagation axis, which is a reasonable approximation due to the small angular scales resulting from the reduced Cherenkov angles from the high-altitude EAS development. Moving away from the propagation direction will result in paths which propagate through both more and less atmosphere, depending on the chosen azimuth angle about the shower axis, giving refraction index profiles $n(L)$ which are increased and decreased, respectively. Thus, we expect larger and smaller time spreads, respectively than what is given using our approximation, which is close to an overall average. 

\begin{figure}[t!]
	\includegraphics[width=\linewidth]{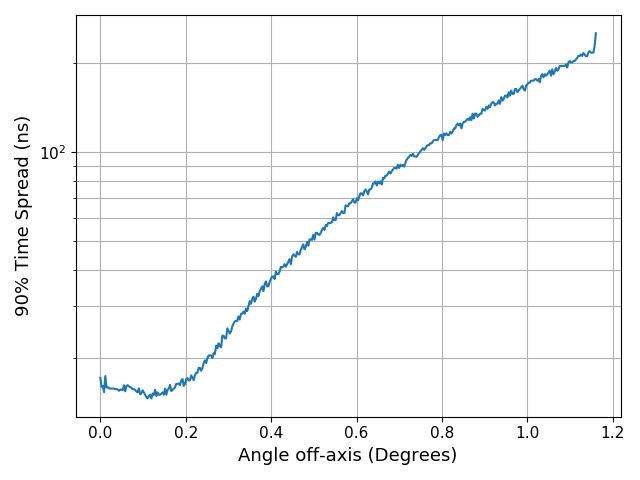}
	\caption{$90\%$ time spread of the arriving Cherenkov photons from a 100~PeV proton shower as observed from 33~km with $\theta_{d}=85^{\circ}$.}
	\label{fig:Example_Timing}
\end{figure}

In conclusion, the arrival time of Cherenkov photons reaching the detection plane is given as the sum $t_{e}(L,E,s)+t_{\gamma}(L,\theta_{\gamma})$. When we spatially histogram the arriving photons, we record also the 90\% spread (between the $5\%$ and $95\%$ percentiles) of their arrival time within each spatial bin and divide the photon density by this time spread to obtain the corresponding photon flux. The $90\%$ time spread of the arriving Cherenkov photons for the flux profile shown in Figure \ref{fig:Mag_Field_Effects} is given in Figure \ref{fig:Example_Timing}. Near axis, the Cherenkov photons arrive within a time window of $\lesssim 20$~ns, while, for observation far away from the shower axis, the time spread increases significantly.

\subsection{Parameter Fitting}
\label{sec:param_fitting}
The calculation of the spatial Cherenkov flux profile uses a significant amount of computational resources, and it is inefficient to perform the calculation every time an event is simulated. Instead, as in \cite{Cummings:2020ycz}, we simulate showers within the geometric parameter space that spans all potentially observable EAS and generate a lookup table of the optical Cherenkov properties to further calculate sensitivities and event rates using a Monte Carlo approach. All simulated showers are generated using a 100~PeV proton primary with the photon yield scaled linearly with the energy of a given EAS.

The first geometric parameter to be sampled is the detector viewing angle $\theta_{d}$. In section \ref{sec:traj_char}, we showed that $\theta_{d}$ is restricted purely by geometry to the ranges $84.2^{\circ}<\theta_{d}<90^{\circ}$ for EUSO-SPB2 and  $67.5^{\circ}<\theta_{d}<70^{\circ}$ for POEMMA. However, we should also take into account the proton interaction length in air and select only $\theta_{d}$ which provide sufficient grammage for first interaction (shower initiation). The average proton-air interaction length as a function of energy is calculated as:
\begin{equation}
\lambda(E) = A m_{N}/\sigma(E)
\label{eq:lambda_proton}
\end{equation}
 
 where $A=14.1$ is the average atomic mass of air, $m_{N}$ is the mass of a nucleon, and $\sigma(E)$ is the proton-air interaction cross section as a function of primary energy. For our purposes here, we use the results published in \cite{Gaisser:2016uoy} to describe the proton-air interaction cross section.
 
 The cumulative probability of interaction within a given depth $X$ is calculated as 
 
 \begin{equation}
 P(E,X) = 1-e^{-X/\lambda(E)} 
 \end{equation}
 
 Using the minimum interaction cross section (corresponding to the largest, and therefore, most restrictive, interaction length) at $E = 1$~PeV, and $P = 0.99$ (that is, 99\% of particle interactions are captured), we calculate an upper bound on $X$ of $300 \, \mathrm{g} \, \mathrm{cm}^{-2}$. For EUSO-SPB2, the geometric range outlined in section \ref{sec:traj_char} satisfies this condition, while, for POEMMA, the allowable angular range shrinks to $67.5^{\circ}<\theta_{d}<68.35^{\circ}$. Both angular ranges are sampled in 50 equally spaced bins. Above the maximum $\theta_{d}$, protons begin to pass through the atmosphere without interacting, leading to their non-detection. We later show that limiting the viewing ranges for POEMMA and EUSO-SPB2 excludes few events, verifying this estimation.

To take into account the proton first interaction depth in the atmosphere, for each $\theta_{d}$ within the detector viewing ranges, we simulate showers with starting depths from $0 \, \mathrm{g} \, \mathrm{cm}^{-2}$ to $280 \, \mathrm{g} \, \mathrm{cm}^{-2}$ in $20 \, \mathrm{g} \, \mathrm{cm}^{-2}$ increments by replacing the longitudinal charged particle profile $N(X)\rightarrow N(X+X_{avg}-X_{0})$. The average first interaction depth of a 100~PeV proton $X_{avg} = 50 \, \mathrm{g} \, \mathrm{cm}^{-2}$ must be included to account for proper shifting of the average charged particle longitudinal profile we use to generate the optical Cherenkov emission. As discussed in \cite{Cummings:2020ycz}, the spatial distribution of the arriving photons as a function of angle off shower axis can be fit with the 5-parameter profile:

\begin{equation}
\Phi_{ch} = 
\begin{cases}
\Phi_{0} & \theta \leq \theta_{ch}\\
\Phi_{0}\Big(\frac{\theta}{\theta_{ch}}\Big)^{-\beta_{F}} & \theta_{ch} \leq \theta \leq \theta_{1}  \\
\Phi_{1}e^{-(\theta-\theta_{1})/\theta_{2}} & \theta \geq \theta_{1}
\end{cases}
\label{1dprof}
\end{equation}

\begin{figure}
	\begin{center}
		\begin{tabular}{cc}
			\centering
			\includegraphics[width=0.5\textwidth]{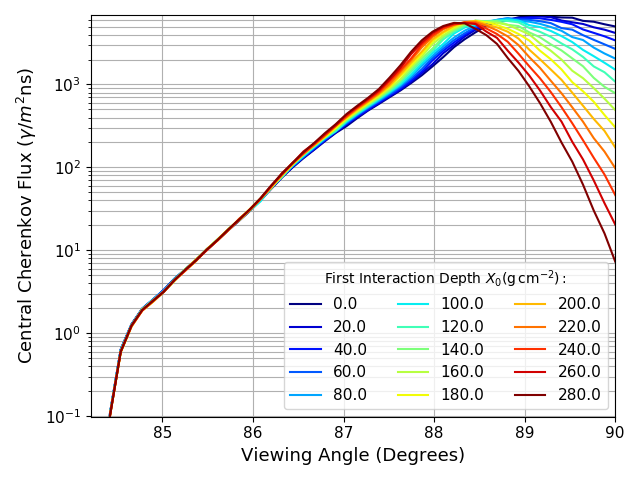} &   \includegraphics[width=0.5\textwidth]{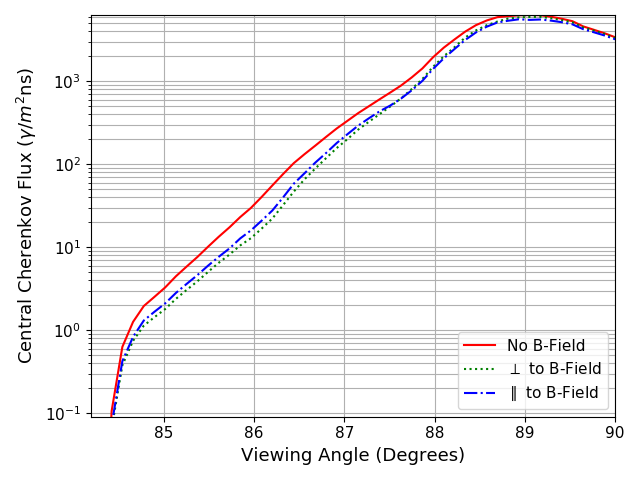} \\
			\includegraphics[width=0.5\textwidth]{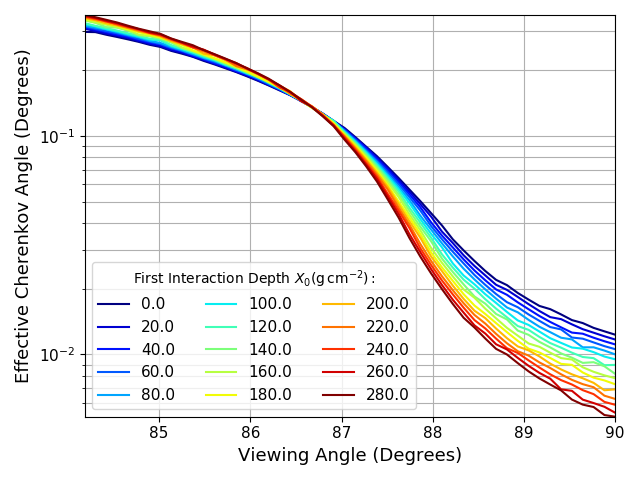} & \includegraphics[width=0.5\textwidth]{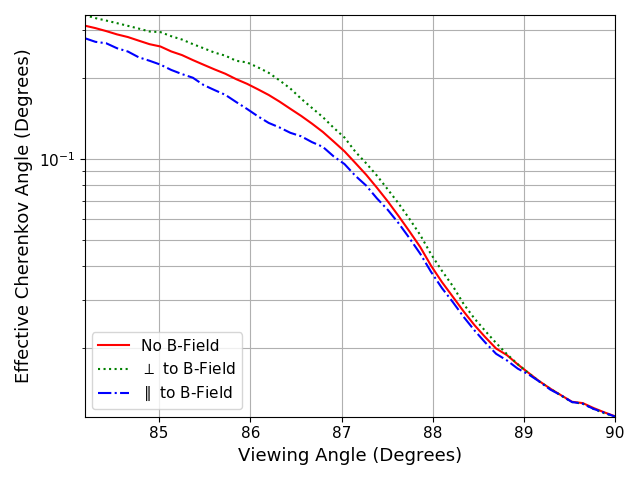} \\
			\includegraphics[width=0.5\textwidth]{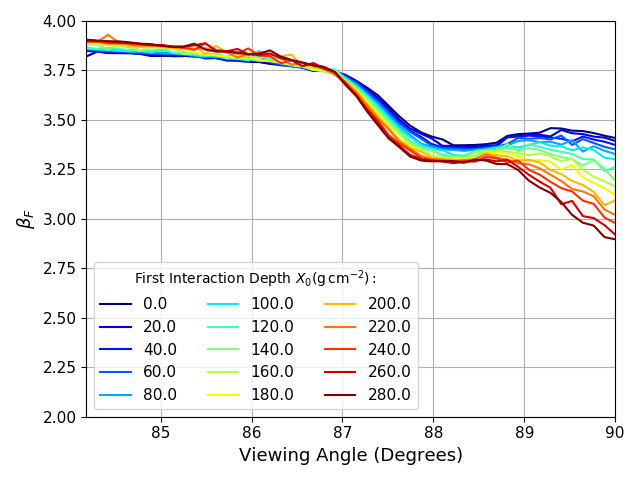} &  \includegraphics[width=0.5\textwidth]{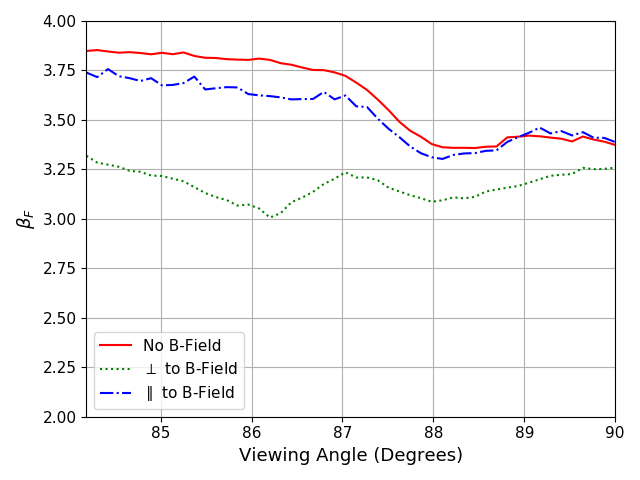} \\
		\end{tabular}
		\end{center}
		\caption{Parameter fits to the Cherenkov spatial distribution from an above-the-limb 100~PeV upward proton shower for different detector viewing angles as observed by a balloon-borne instrument at 33~km altitude. Plots are central flux  $\Phi_{0}$ [upper panel], central width $\theta_{ch}$ [middle panel], and power law scale $\beta_{F}$ [lower panel] using the 5 parameter fit model described in equation \ref{1dprof}. The left panels show the effects of changing the first interaction depth of the shower, while the right panels show the effect of the geomagnetic separation of the positrons and electrons in the shower, using a representative starting depth of $40 \, \mathrm{g} \, \mathrm{cm}^{-2}$.}
		\label{fig:SPB_Params}
	
\end{figure}

\begin{figure}
	\begin{center}
		\begin{tabular}{cc}
			\centering
			\includegraphics[width=0.5\textwidth]{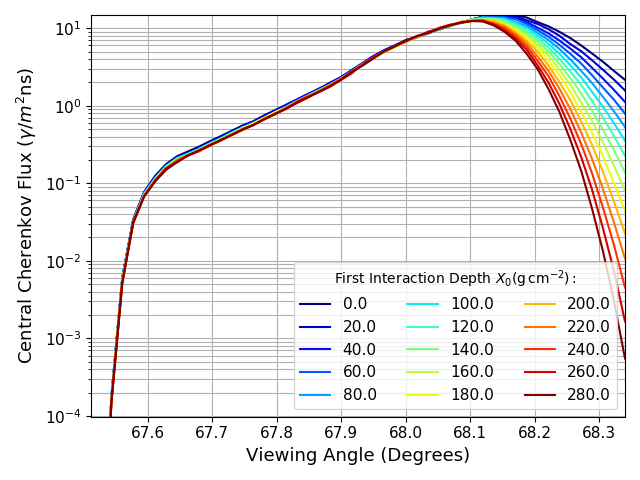} &   \includegraphics[width=0.5\textwidth]{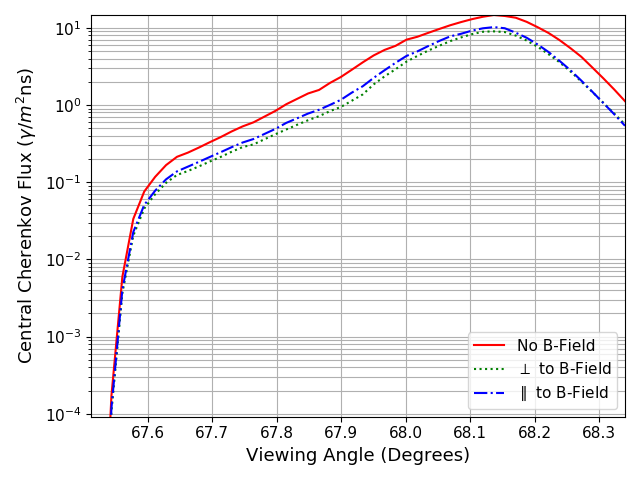} \\
			\includegraphics[width=0.5\textwidth]{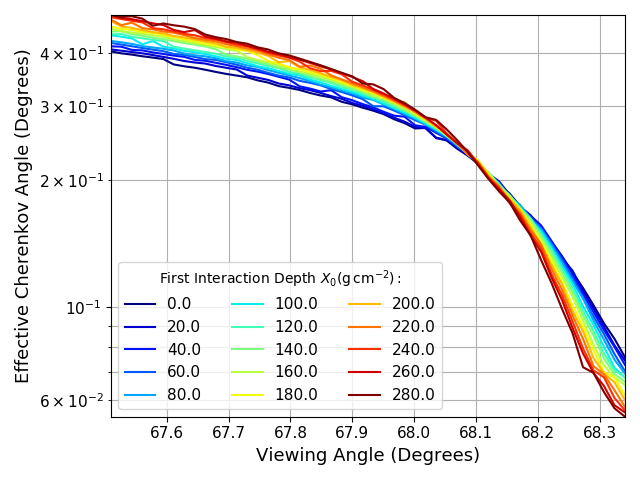} & \includegraphics[width=0.5\textwidth]{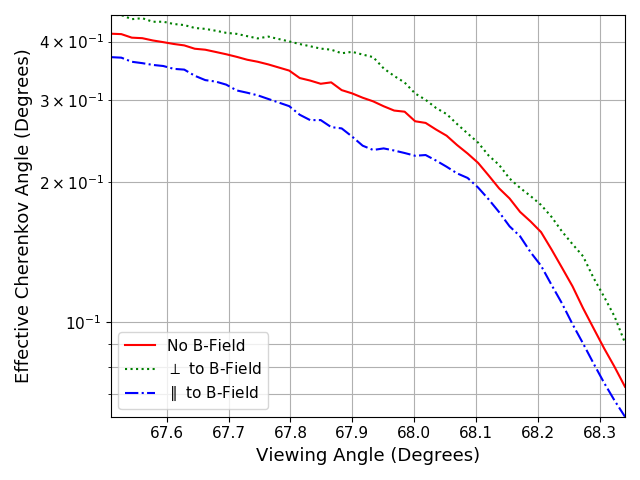} \\
			\includegraphics[width=0.5\textwidth]{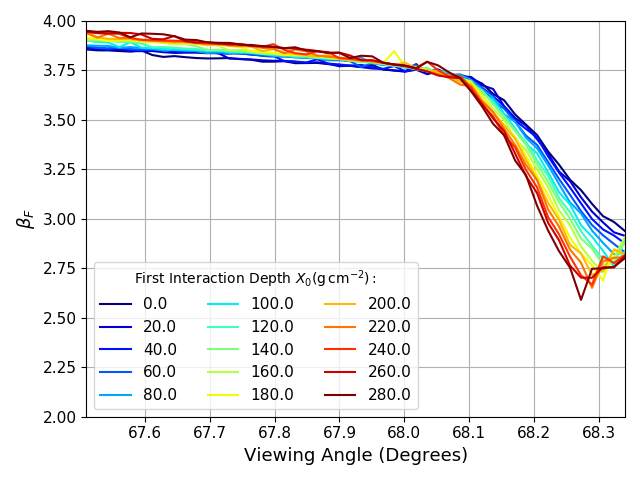} &  \includegraphics[width=0.5\textwidth]{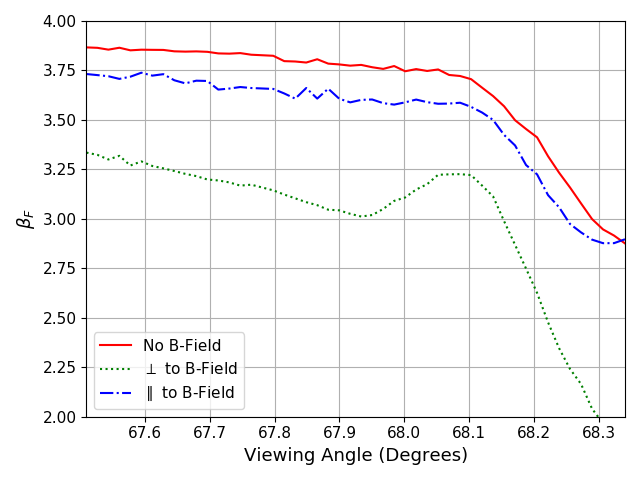} \\
		\end{tabular}
	\end{center}
	\caption{Parameter fits to the Cherenkov spatial distribution from an above-the-limb 100~PeV upward proton shower for different detector viewing angles as observed by a satellite-based instrument at 525~km altitude. Plots are central flux  $\Phi_{0}$ [upper panel], central width $\theta_{ch}$ [middle panel], and power law scale $\beta_{F}$ [lower panel] using the 5 parameter fit model described in equation \ref{1dprof}. The left panels show the effects of changing the first interaction depth of the shower, while the right panels show the effect of the geomagnetic separation of the positrons and electrons in the shower, using a representative starting depth of $40 \, \mathrm{g} \, \mathrm{cm}^{-2}$.}
	\label{fig:POEMMA_Params}
	
\end{figure}

This model assumes a uniform optical Cherenkov photon flux $\Phi_{0}$ within an angle $\theta_{ch}$, a $\beta_{F}$ power-law fall off within an angle $\theta_{1}$ and, for larger angles, an exponential fall off with scale $\theta_{2}$. The flux $\Phi_{1}$ is calculated such that the function remains continuous between the last 2 regions, and is not a free parameter. All angles are given with respect to the shower axis as measured from the top of the atmosphere (the point where the shower enters the atmosphere, not to be confused with the first interaction point). Example fits to the spatial distribution of the arriving Cherenkov photon flux are shown in Figure \ref{fig:Mag_Field_Effects}. These fits represent an overall average near the shower axis, taking into account the Cherenkov ``horn'' features, and provide an excellent fitting of the tails of the distributions even out to large viewing angles. For each simulated shower, we perform the fit to (i) the spatial distribution without a magnetic field (ii) the spatial distribution along the axis of the magnetic field (iii) the spatial distribution along the axis perpendicular to the magnetic field. This allows for a range of values that a shower is likely to exhibit during propagation. In Figures \ref{fig:SPB_Params} and \ref{fig:POEMMA_Params}, we respectively plot the three parameters that dominate the behavior of the spatial distribution of the optical Cherenkov flux close to shower axis ($\Phi_{0}, \theta_{ch}, \beta_{F}$) as a function of detector viewing angle $\theta_{d}$ for the altitudes corresponding to the EUSO-SPB2 and POEMMA instruments. In the left panels of the two figures, we show the effect of changing the first interaction depth $X_{0}$ on the parameter fits, while in the right panels, we show the effect of the application of the geomagnetic field as described above.

For small detector viewing angles (events viewed at small angles from the Earth limb), the intensity of the signal is minimized, due to the strong atmospheric extinction of the optical Cherenkov emission. Specifically, trajectories with points of lowest approach less than roughly 5~km (see Figures \ref{fig:grammage} and \ref{fig:Optical_Depth}) experience aerosol dominated scattering and long path lengths through the atmosphere, thereby providing for strong attenuation of the arriving photons. Consequently, as the detector viewing angle increases, the signal intensity increases due to the decreased atmospheric attenuation. This trend continues until a certain viewing angle, where the intensity begins to decrease for two reasons: (i) the path length through the atmosphere becomes too thin to support significant development of the EAS and (ii) the Cherenkov threshold becomes very high ($\sim 1$~GeV) such that only the highest energy electrons can contribute to Cherenkov photon generation. This trend occurs earlier for events with larger first interaction depths, as earlier portions of the shower are being sampled, where there is inherently less charged particle content. 

With increasing viewing angle, the spatial distribution of the optical Cherenkov emission becomes more narrow for two key reasons: (i) due to the decreased index of refraction at high altitudes, the local Cherenkov angle becomes smaller, thereby focusing the Cherenkov photons onto a tighter cone, which again is a stronger effect for events with large first interaction depths, as they produce their maximum photon content at higher altitudes (ii) the energy threshold of Cherenkov generation is increased at higher altitudes and therefore electrons with characteristically larger energies (which remain closer to the shower axis) generate the observable emission. From this, we expect to observe steeper tails of the distribution with increasing detector viewing angle. However, taking into account the electron lateral distribution, which also becomes more prevalent at high altitudes, as discussed in \cite{Cummings:2020ycz}, we actually see a broadening of the tails of the distribution with increasing viewing angle. In a balloon-borne framework, when the detector viewing angle is large enough ($\sim 4^{\circ}$ above the limb), the instrument sits inside active shower development, resulting in a very spatially contained (nearly exponential), bright signal.

We also note the effect of the geomagnetic field, which is to spread the signal effectively along the axis perpendicular to the field. The central optical Cherenkov intensity of the distributions from showers affected by the geomagnetic field is smaller than the unmodified showers up to a maximum of a factor of 2, following the right panels of Figures \ref{fig:SPB_Params} and \ref{fig:POEMMA_Params}. The effective Cherenkov angle of the distribution is slightly larger along the axis perpendicular to the magnetic field and smaller along the axis parallel to the magnetic field compared to the unmodified distribution. Similarly, the tails of the distribution are less steep for any shower modified by the Earth magnetic field than the tails from unmodified showers, especially if measuring perpendicular to the applied field (as expected). Thus, showers affected by the geomagnetic field may be less bright directly on axis, but can have significantly higher photon yields further off axis, making them less observable at low primary energies and more observable at the highest energies.  

\onecolumngrid

\newpage
\twocolumngrid

\begin{figure}[t!]
\includegraphics[width=\linewidth]{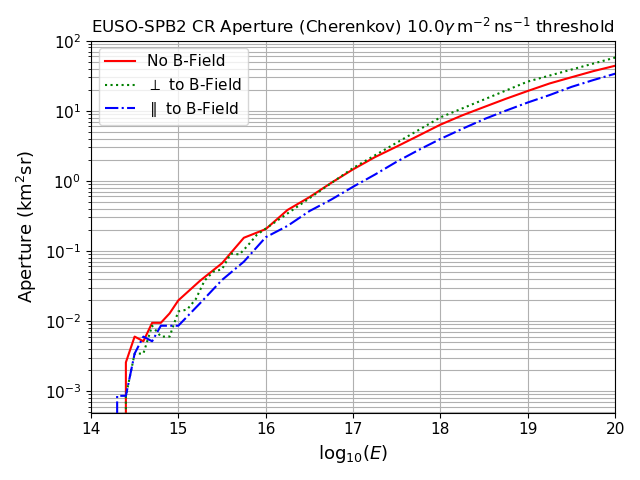}
\includegraphics[width=\linewidth]{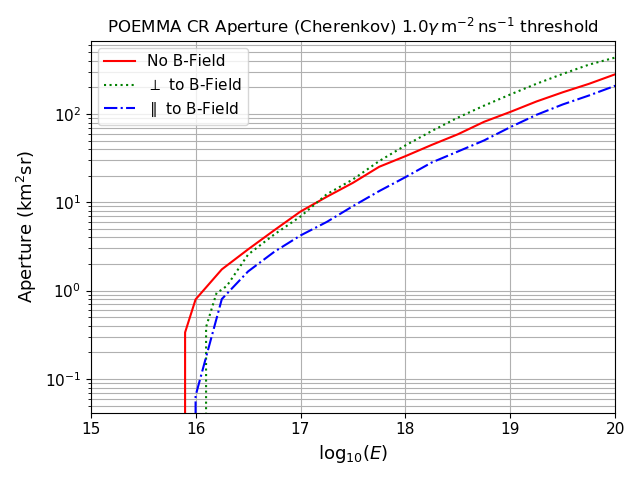}
\caption{Geometric aperture to above-the-limb cosmic rays as a function of primary energy for the EUSO-SPB2 [upper panel] and POEMMA [lower panel] detectors.}
\label{fig:Apertures}
\end{figure}

\begin{figure}[t!]
	\includegraphics[width=\linewidth]{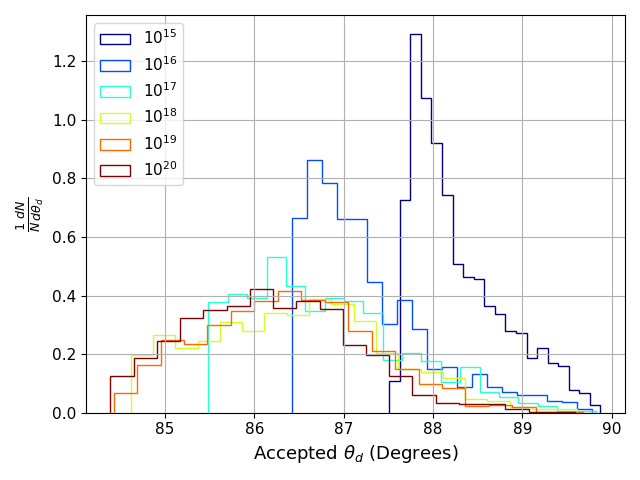}
	\includegraphics[width=\linewidth]{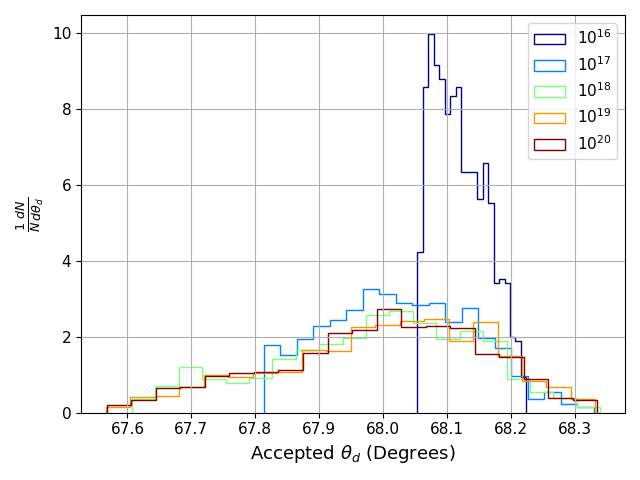}
	\caption{Normalized distribution of arrival angle $\theta_{d}$ for accepted above-the-limb cosmic rays for different primary energies as measured with the EUSO-SPB2 instrument [upper panel] and POEMMA instrument [lower panel].}
	\label{fig:above_limb_arrival_angle_distribution}
\end{figure}

\section{Aperture and event rate}
\label{sec:aper_sens}

For the Earth-skimming neutrino events analyzed in \cite{Cummings:2020ycz}, a semi-analytical estimate was used to determine the geometric aperture and sensitivity, where a Monte Carlo methodology was used only to estimate the average behavior of the EAS properties for use in the simulation. This was due, in part, to the sheer number of events which needed to be simulated (correspondingly, the large amount of computation time) in order to calculate an accurate figure, properly sampling all the relevant distributions involved. When simulating cosmic ray events from above the limb, we do not have these restrictions, as proton induced EAS vary significantly mainly by the first interaction depth which decreases with increasing energy (here we do not consider the Landau-Migdal-Pomeranchuk (LPM) effect \cite{Cillis_1999} or $\pi^{0}$ interactions, which for $z>20$~km become relevant for energies greater than $3 \times 10^{18}$~eV and $7 \times 10^{19}$~eV, respectively) \cite{cummings2019complete}. Additionally, as observed in Figures \ref{fig:SPB_Params} and \ref{fig:POEMMA_Params}, the intensity and the angular scales of the Cherenkov emission from above-the-limb cosmic ray EAS vary rapidly with detector viewing angle (on scales smaller than the effective Cherenkov angle of the distribution), making an analytical estimate unreliable. For these reasons, in the present computation scheme we utilize a more realized Monte Carlo methodology.

In the Earth-centered coordinate system shown in Figure \ref{fig:Geometry}, the detector is positioned at the cartesian coordinates (0, 0, $R_{E}+h$), where $R_{E}$ is the Earth radius and $h$ the detector altitude above ground (33~km for EUSO-SPB2, 525~km for POEMMA). The starting point of the shower is sampled isotropically on the top of Earth's atmosphere, namely with radius $R_{E}+z_{atm}$, zenith angle sampled uniformly in cos$\theta_{E}$ within the detector viewing range and azimuth $\phi_E$ sampled uniformly between (0, $2\pi$). 

The trajectory of the shower must also then be sampled isotropically. To do this, we sample the shower zenith in its local frame uniformly in $\mathrm{cos}^{2}\theta_{tr}$ to take into account the probability that a trajectory will lie within an angular bin $d\theta_{tr}$ on a planar surface. We then sample the azimuth  of the trajectory $\phi_{tr}$ uniformly on (0, $2\pi$). Converting to Cartesian coordinates, we then calculate the angle between this trajectory and the detector viewing direction to the sampled shower starting point.

The first interaction depth is sampled from an exponential distribution with mean $\lambda$, given in equation \ref{eq:lambda_proton}.  Using the sampled shower trajectory and first interaction depth, as well as the lookup table of fit parameters of the Cherenkov spatial distribution for a 100~PeV primary shower, we then determine the intensity of Cherenkov photons at the detector from a given shower, scaling the profile by $E/100$~PeV for different primary energies. 

If the intensity exceeds the threshold determined by the optics and electronics of the instrument, the event is accepted. The choice of the detectable photon threshold for the two instruments is $20 \,  \gamma/\mathrm{m}^{2}$ for POEMMA and $200 \, \gamma/\mathrm{m}^{2}$ for EUSO-SPB2, taking into consideration the effects of the dark-sky background, as outlined in \cite{Cummings:2020ycz, Reno:2019jtr}. We then divide this calculated threshold by the minimum electronics integration time, which, for this study, we consider to be 20~ns. The spread in arrival times of the observed Cherenkov photons exceeds roughly 20~ns regardless of how far the photons are viewed from the shower axis, so we are justified in comparing the flux of arriving photons with the calculated threshold. That is, we do not need to consider dim events which arrive with very short time scales within our analysis. For every primary cosmic ray energy simulated, we simulate $10^{5}$ shower starting points and $10^{5}$ shower trajectories for each sampled starting point. The geometric aperture is then calculated as:

\begin{equation}
\begin{split}
\langle A\Omega \rangle (E) &= \pi S  \frac{N_{accepted}(E)}{N_{simulated}(E)} \\
S &= \int_{0}^{2\pi} \int_{\theta_{E1}}^{\theta_{E2}} (R_{E}+z_{atm})^{2} \mathrm{sin}\theta_{E} d\theta_{E} d\phi
\end{split}
\end{equation}

where $S$ is the area of the disk visible to the instrument \cite{CRANNELL1971179, SULLIVAN19715}. To place bounds on the possible effects of including the Earth geomagnetic field, we perform this calculation using the fits to the Cherenkov spatial distribution under the application of the magnetic field as described in section \ref{sec:geofield}, measured along the axes parallel and perpendicular to the applied field, as well as by applying no field. By doing so, we assume that all showers propagating within the geomagnetic field experience maximal (perpendicular to $B$) and minimal (parallel to $B$) deflection of the electrons and positrons generating the optical Cherenkov emission. In Figure \ref{fig:Apertures}, we plot the geometric apertures of the EUSO-SPB2 (upper panel) and POEMMA (lower panel) experiments to the above-the-limb cosmic rays.

Figure \ref{fig:Apertures} demonstrates that, as expected, the EUSO-SPB2 instrument has increased sensitivity at energies below 10~PeV due to being closer to the shower development. At higher energies, the sensitivity of the POEMMA instrument begins to dominate, becoming roughly an order of magnitude larger than that of EUSO-SPB2 at $10^{20}$~eV. We also note that the effect of including the Earth geomagnetic field is modest. Whether all cosmic ray events are measured completely along the axis parallel or perpendicular to the magnetic field, the geometric acceptances remain within a factor of two of the case with no applied field. Additionally, as the events measured perpendicular to the magnetic  field provide an increase in geometric aperture with respect to the unaffected case (due to the strengthening of the tails of the distribution), and those measured parallel result in a decrease, an average measurement will occur in-between these bounds and minimize the effect. Although, it is worth noting that, because the central intensity of the optical Cherenkov spatial distribution is decreased due to the geomagnetic effect, the energy threshold is slightly increased for events affected by a magnetic field with respect to the standard (no magnetic field) case.

\begin{figure}[t!]
	\includegraphics[width=\linewidth]{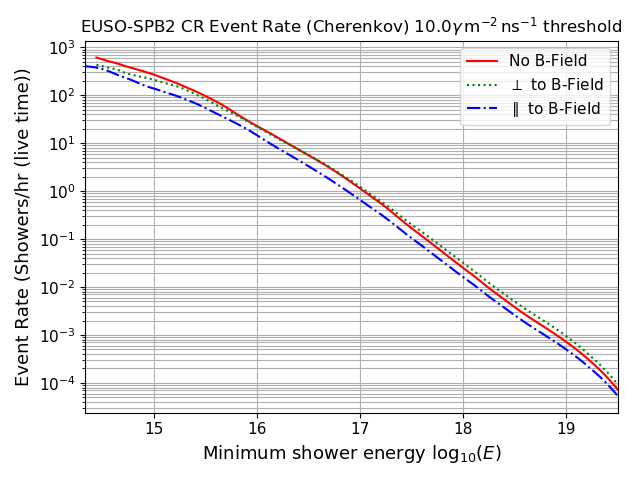}
	\includegraphics[width=\linewidth]{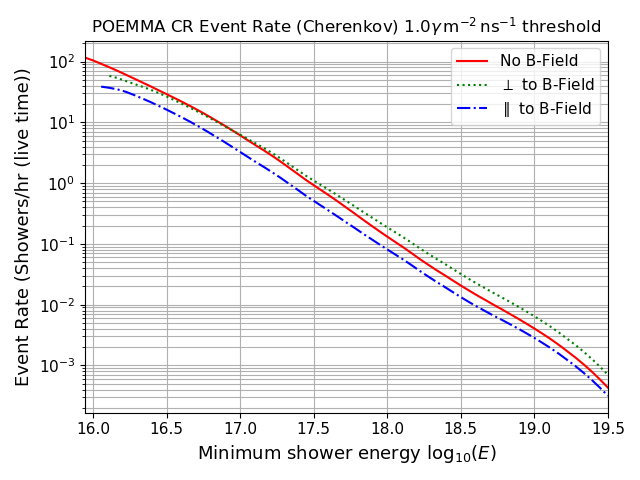}
	\caption{Integrated expected event rate (events measured above given energy $E$) for above-the-limb UHECR events for the EUSO-SPB2 [upper panel] and POEMMA [lower panel] instruments. Event rate is given per hour of live time (instrument duty cycle for each not taken into account).}
	\label{fig:Event_Rates}
\end{figure}

We also plot the normalized distribution of arrival angles of the accepted events in figure \ref{fig:above_limb_arrival_angle_distribution}. Figure \ref{fig:above_limb_arrival_angle_distribution} shows that the chosen range for $\theta_{d}$ as outlined in section \ref{sec:param_fitting} is well motivated, with few events being accepted above $90^{\circ}$ and $68.35^{\circ}$ for EUSO-SPB2 and POEMMA, respectively. We also observe that with increasing primary energy, more events are accepted closer to the Earth limb, due to the brightening of signals which are able to compete with the heavy atmospheric extinction, as expected. Overall, the majority of the above-the-limb cosmic rays are accepted with angles $\theta_{d}\sim 86.5^{\circ}$ and $\theta_{d}\sim 68.1^{\circ}$ for EUSO-SPB2 and POEMMA, respectively. Comparing with figure \ref{fig:grammage}, both of these correspond to total path lengths of $\sim 2000 \, \mathrm{g} \, \mathrm{cm}^{-2}$. As a point of reference, the thickness of the vertical Earth atmosphere is $1030 \, \mathrm{g} \, \mathrm{cm}^{-2}$. 

Given an expected flux of cosmic rays $\Phi_{CR}(E)$ and the geometric aperture $\langle A\Omega \rangle (E)$, we calculate the estimated event rate above an energy $E$ as:

\begin{equation}
N = \int \int_{E}^{\infty} \langle A\Omega \rangle (E) \Phi _{CR}(E) d E dt
\label{Nevents}
\end{equation}

where we specifically define the event rate here to be the number of expected events above a given energy $E$. Concerning the choice of the cosmic ray flux $\Phi_{CR}(E)$, we use the combined data of the all particle energy spectra from the Tibet-ASg, KASCADE-Grande, and Pierre-Auger experiments across the range $(10^{14}~\mathrm{eV},10^{20}~\mathrm{eV})$ as given in \cite{Gaisser:2016uoy}. In figure \ref{fig:Event_Rates}, we plot the estimated event rate per hour of live time of above-the-limb cosmic rays as a function of minimum energy in the case of EUSO-SPB2 and POEMMA. Figure \ref{fig:Event_Rates} demonstrates that, for both orbital and sub-orbital Cherenkov telescopes, the expected event rate is very large, being hundreds of events per hour of live time above energies of less than a PeV in the sub-orbital case and 10~PeV in the orbital case.

\section{Conclusions}
\label{sec:conc}

In this paper, we have extended the computation scheme developed in \cite{Cummings:2020ycz} to also model the observation of the cosmic ray EAS which arrive with trajectories from above Earth's limb and calculate their expected event rate for the EUSO-SPB2 and POEMMA instruments. Cosmic rays can deposit much of their primary energy into showering products, resulting in extremely bright optical Cherenkov signals.

In section \ref{sec:traj_char}, we discussed the characteristic altitudes of the shower development for above-the-limb EAS which occur above 20~km, where the atmosphere is rarified, leading to signals with unique properties. Specifically, we showed that the required thresholds for optical Cherenkov emission are increased, with smaller local Cherenkov angles, while the atmospheric transmission can be greatly decreased with respect to upward going (below the limb) EAS. These combined effects result in bright signals which are focused close to the shower propagation axis.

As these events can be extremely bright, even for large angles off shower axis, it was necessary to consider also the time spread of arriving photons at the plane of detection, which can increase up to a few microseconds when measured far off axis, much greater than the typical $10-20$~ns integration time of the Cherenkov telescope designs being investigated. This fact implies a reduction of the estimated geometric aperture to above-the-limb cosmic ray events, with the larger effect at the highest energies, where the exponential tails of the optical Cherenkov spatial distribution become relevant.

Additionally, for shower development within a rarified atmosphere (high altitudes), the distance scale corresponding to a radiation length is much longer than that at low atmospheric altitudes, allowing for more significant geomagnetic deflection of electrons and positrons. To consider the effects of the geomagnetic field, we took the approach of applying a relatively large (50~$\mu$T) field perpendicular to the shower propagation direction, and measured the flux profile of arriving Cherenkov photons along the axes perpendicular and parallel to the magnetic field compared with the profile of unaffected showers (symmetric about the shower axis). We demonstrated that the effect of applying a magnetic field to the developing EAS is to spread the optical Cherenkov photons within the effective Cherenkov angle away from shower axis along the axis perpendicular to the magnetic field, thereby reducing the central intensity, but increasing the intensity within the tails of the distribution. This approach provided an upper and lower bound on the effect of magnetic deflection, showing that, ultimately, it is a modest, factor of $\sim 2$, effect on the Cherenkov intensity for a specific EAS energy and trajectory.

Using a Monte Carlo methodology, we showed that the estimated event rate of (above-the-limb) cosmic rays for the EUSO-SPB2 and POEMMA instruments can be very high. Specifically, as follows from Figures \ref{fig:Apertures} and \ref{fig:Event_Rates}, we see that both instruments have the capability to observe potentially hundreds of events per hour of live time above energies of 300~TeV and 10~PeV for sub-orbital and orbital observation schemes, respectively.


The properties of the optical Cherenkov emission from the above-the-limb cosmic rays are extremely similar to those of the neutrino events of comparable energy in wavelength, arrival angle, and arrival time distributions despite the development at high altitudes. Taking this information together with the huge event rates presented in figure \ref{fig:Event_Rates}, the above-the-limb cosmic rays represent a guaranteed in-flight test source for both orbital and sub-orbital optical Cherenkov telescopes. While simulation studies have predicted the rate of upward-moving EAS sourced from Earth-interacting neutrino events to be small for cosmogenic flux assumptions \cite{Cummings:2020ycz,Reno:2019jtr}, above-the-limb cosmic rays are plentiful, allowing for validation of the optical Cherenkov detection technique for upward going EAS. Observation of the above-the-limb cosmic rays will also help to quantify optical performance, trigger algorithms, and other detector properties such that the Cherenkov telescope can be optimized for neutrino detection.

 Atmospheric refraction near the Earth limb can be up to $\sim 1^{\circ}$ \cite{Chu:83}. As shown in Figure \ref{fig:above_limb_arrival_angle_distribution}, for above-the-limb cosmic rays with $E>10^{19}$~eV, there is limited acceptance very close to the limb. As such, it will likely be necessary to correspondingly restrict neutrino searches to at least $1^{\circ}$ below the limb to avoid false positive signatures triggered by atmospherically refracted signals from cosmic rays \cite{Venters:2019xwi}.

The above-the-limb cosmic ray events are ideal candidates for energy and directional reconstruction for four reasons: (i) the signal intensity is approximately linearly proportional to the primary cosmic ray energy (ii) the angular acceptance is energy dependent, see Figure \ref{fig:above_limb_arrival_angle_distribution} (iii) the overall angular scales are small due to the high altitude development, leading to exceptional angular resolution of the trajectory of the EAS (iv) the high number of events allows for large statistical groupings. The techniques developed to identify and reconstruct these events from eventual flight data will also be used on neutrino candidate events and will therefore need to be intensively studied and refined such that they are well understood and optimized for neutrino detection.

Mass composition measurements of the above-the-limb cosmic rays using the optical Cherenkov detection technique might also be performed by mirroring the observation strategies utilized for ground based Cherenkov telescopes, mainly: (i) The observation and discrimination of the Cherenkov emission generated by muons within the EAS, which helps to clarify the nature of the primary particle \cite{Neronov_2016,krolik2019cherenkov} and (ii) Multiple observations within the effective Cherenkov angle of the resulting emission, which allows for an estimation of $X_\mathrm{max}$ on an EAS-by-EAS basis \cite{Tunka1,Tunka2,krizmanic2013nonimaging,krizmanic2013modeling,2015ICRC...34..635B, 2015ICRC...34..562K, 2017ICRC...35..415B,Omura:2020dzl,Bergman:2019f0}. While both of these methodologies deserve a detailed study to examine their feasibility in high altitude observations of the above-the-limb cosmic rays, we note that, in principle, a single imaging Cherenkov telescope such as that of EUSO-SPB2 or POEMMA, is capable of aiming to perform the first analysis, while for the second, an array of future orbital Cherenkov instruments is required to image the $\mathcal{O}(10~\mathrm{km})$ Cherenkov light pool. Such strategies may potentially allow for composition measurements of the cosmic ray flux across a wide range of primary energies.

The work and results presented in this paper  provide an initial step in the assessment of the full scientific potential of above-the-limb cosmic ray optical Cherenkov light measurements.

 
\bibliographystyle{apsrev4-1}
\bibliography{biblio.bib}

\begin{thebibliography}{54}%
\makeatletter
\providecommand \@ifxundefined [1]{%
 \@ifx{#1\undefined}
}%
\providecommand \@ifnum [1]{%
 \ifnum #1\expandafter \@firstoftwo
 \else \expandafter \@secondoftwo
 \fi
}%
\providecommand \@ifx [1]{%
 \ifx #1\expandafter \@firstoftwo
 \else \expandafter \@secondoftwo
 \fi
}%
\providecommand \natexlab [1]{#1}%
\providecommand \enquote  [1]{``#1''}%
\providecommand \bibnamefont  [1]{#1}%
\providecommand \bibfnamefont [1]{#1}%
\providecommand \citenamefont [1]{#1}%
\providecommand \href@noop [0]{\@secondoftwo}%
\providecommand \href [0]{\begingroup \@sanitize@url \@href}%
\providecommand \@href[1]{\@@startlink{#1}\@@href}%
\providecommand \@@href[1]{\endgroup#1\@@endlink}%
\providecommand \@sanitize@url [0]{\catcode `\\12\catcode `\$12\catcode
  `\&12\catcode `\#12\catcode `\^12\catcode `\_12\catcode `\%12\relax}%
\providecommand \@@startlink[1]{}%
\providecommand \@@endlink[0]{}%
\providecommand \url  [0]{\begingroup\@sanitize@url \@url }%
\providecommand \@url [1]{\endgroup\@href {#1}{\urlprefix }}%
\providecommand \urlprefix  [0]{URL }%
\providecommand \Eprint [0]{\href }%
\providecommand \doibase [0]{http://dx.doi.org/}%
\providecommand \selectlanguage [0]{\@gobble}%
\providecommand \bibinfo  [0]{\@secondoftwo}%
\providecommand \bibfield  [0]{\@secondoftwo}%
\providecommand \translation [1]{[#1]}%
\providecommand \BibitemOpen [0]{}%
\providecommand \bibitemStop [0]{}%
\providecommand \bibitemNoStop [0]{.\EOS\space}%
\providecommand \EOS [0]{\spacefactor3000\relax}%
\providecommand \BibitemShut  [1]{\csname bibitem#1\endcsname}%
\let\auto@bib@innerbib\@empty
\bibitem [{\citenamefont {Cummings}\ \emph {et~al.}(2021)\citenamefont
  {Cummings}, \citenamefont {Aloisio},\ and\ \citenamefont
  {Krizmanic}}]{Cummings:2020ycz}%
  \BibitemOpen
  \bibfield  {author} {\bibinfo {author} {\bibfnamefont {A.}~\bibnamefont
  {Cummings}}, \bibinfo {author} {\bibfnamefont {R.}~\bibnamefont {Aloisio}}, \
  and\ \bibinfo {author} {\bibfnamefont {J.}~\bibnamefont {Krizmanic}},\ }\href
  {\doibase 10.1103/physrevd.103.043017} {\bibfield  {journal} {\bibinfo
  {journal} {Physical Review D}\ }\textbf {\bibinfo {volume} {103}} (\bibinfo
  {year} {2021}),\ 10.1103/physrevd.103.043017}\BibitemShut {NoStop}%
\bibitem [{\citenamefont {Reno}\ \emph {et~al.}(2019)\citenamefont {Reno},
  \citenamefont {Krizmanic},\ and\ \citenamefont {Venters}}]{Reno:2019jtr}%
  \BibitemOpen
  \bibfield  {author} {\bibinfo {author} {\bibfnamefont {M.~H.}\ \bibnamefont
  {Reno}}, \bibinfo {author} {\bibfnamefont {J.~F.}\ \bibnamefont {Krizmanic}},
  \ and\ \bibinfo {author} {\bibfnamefont {T.~M.}\ \bibnamefont {Venters}},\
  }\href {\doibase 10.1103/PhysRevD.100.063010} {\bibfield  {journal} {\bibinfo
   {journal} {Phys. Rev.}\ }\textbf {\bibinfo {volume} {D100}},\ \bibinfo
  {pages} {063010} (\bibinfo {year} {2019})},\ \Eprint
  {http://arxiv.org/abs/1902.11287} {arXiv:1902.11287 [astro-ph.HE]}
  \BibitemShut {NoStop}%
\bibitem [{\citenamefont {Aloiso}\ \emph {et~al.}(2018)\citenamefont {Aloiso},
  \citenamefont {Coccia},\ and\ \citenamefont {Vissani}}]{Aloiso:2018hbl}%
  \BibitemOpen
  \bibinfo {editor} {\bibfnamefont {R.}~\bibnamefont {Aloiso}}, \bibinfo
  {editor} {\bibfnamefont {E.}~\bibnamefont {Coccia}}, \ and\ \bibinfo {editor}
  {\bibfnamefont {F.}~\bibnamefont {Vissani}},\ eds.,\ \href {\doibase
  10.1007/978-3-319-65425-6} {\emph {\bibinfo {title} {{Multiple Messengers and
  Challenges in Astroparticle Physics}}}}\ (\bibinfo  {publisher} {Springer},\
  \bibinfo {address} {Cham},\ \bibinfo {year} {2018})\BibitemShut {NoStop}%
\bibitem [{\citenamefont {Wiencke}\ and\ \citenamefont
  {Olinto}(2020)}]{Wiencke:2019vke}%
  \BibitemOpen
  \bibfield  {author} {\bibinfo {author} {\bibfnamefont {L.}~\bibnamefont
  {Wiencke}}\ and\ \bibinfo {author} {\bibfnamefont {A.}~\bibnamefont {Olinto}}
  (\bibinfo {collaboration} {JEM-EUSO, POEMMA}),\ }\href {\doibase
  10.22323/1.358.0466} {\bibfield  {journal} {\bibinfo  {journal} {PoS}\
  }\textbf {\bibinfo {volume} {ICRC2019}},\ \bibinfo {pages} {466} (\bibinfo
  {year} {2020})},\ \Eprint {http://arxiv.org/abs/1909.12835} {arXiv:1909.12835
  [astro-ph.IM]} \BibitemShut {NoStop}%
\bibitem [{\citenamefont {Olinto}\ \emph {et~al.}(2020)\citenamefont {Olinto}
  \emph {et~al.}}]{Olinto:2020oky}%
  \BibitemOpen
  \bibfield  {author} {\bibinfo {author} {\bibfnamefont {A.}~\bibnamefont
  {Olinto}} \emph {et~al.},\ }\href@noop {} {\  (\bibinfo {year} {2020})},\
  \Eprint {http://arxiv.org/abs/2012.07945} {arXiv:2012.07945 [astro-ph.IM]}
  \BibitemShut {NoStop}%
\bibitem [{\citenamefont {Olinto}\ \emph {et~al.}(2019)\citenamefont {Olinto}
  \emph {et~al.}}]{Olinto:2019mjh}%
  \BibitemOpen
  \bibfield  {author} {\bibinfo {author} {\bibfnamefont {A.}~\bibnamefont
  {Olinto}} \emph {et~al.},\ }\href@noop {} {\  (\bibinfo {year} {2019})},\
  \Eprint {http://arxiv.org/abs/1907.06217} {arXiv:1907.06217 [astro-ph.HE]}
  \BibitemShut {NoStop}%
\bibitem [{\citenamefont {Olinto}\ \emph {et~al.}(2018)\citenamefont {Olinto}
  \emph {et~al.}}]{Olinto:2017xbi}%
  \BibitemOpen
  \bibfield  {author} {\bibinfo {author} {\bibfnamefont {A.~V.}\ \bibnamefont
  {Olinto}} \emph {et~al.},\ }\href {\doibase 10.22323/1.301.0542} {\bibfield
  {journal} {\bibinfo  {journal} {PoS}\ }\textbf {\bibinfo {volume}
  {ICRC2017}},\ \bibinfo {pages} {542} (\bibinfo {year} {2018})},\ \bibinfo
  {note} {[35,542(2017)]},\ \Eprint {http://arxiv.org/abs/1708.07599}
  {arXiv:1708.07599 [astro-ph.IM]} \BibitemShut {NoStop}%
\bibitem [{\citenamefont {Barwick}\ \emph {et~al.}(2006)\citenamefont {Barwick}
  \emph {et~al.}}]{Barwick:2005hn}%
  \BibitemOpen
  \bibfield  {author} {\bibinfo {author} {\bibfnamefont {S.~W.}\ \bibnamefont
  {Barwick}} \emph {et~al.} (\bibinfo {collaboration} {ANITA}),\ }\href
  {\doibase 10.1103/PhysRevLett.96.171101} {\bibfield  {journal} {\bibinfo
  {journal} {Phys. Rev. Lett.}\ }\textbf {\bibinfo {volume} {96}},\ \bibinfo
  {pages} {171101} (\bibinfo {year} {2006})},\ \Eprint
  {http://arxiv.org/abs/astro-ph/0512265} {arXiv:astro-ph/0512265 [astro-ph]}
  \BibitemShut {NoStop}%
\bibitem [{\citenamefont {Gorham}\ \emph {et~al.}(2009)\citenamefont {Gorham}
  \emph {et~al.}}]{2009APh....32...10A}%
  \BibitemOpen
  \bibfield  {author} {\bibinfo {author} {\bibfnamefont {P.~W.}\ \bibnamefont
  {Gorham}} \emph {et~al.},\ }\href {\doibase
  10.1016/j.astropartphys.2009.05.003} {\bibfield  {journal} {\bibinfo
  {journal} {Astroparticle Physics}\ }\textbf {\bibinfo {volume} {32}},\
  \bibinfo {pages} {10} (\bibinfo {year} {2009})},\ \Eprint
  {http://arxiv.org/abs/0812.1920} {arXiv:0812.1920 [astro-ph]} \BibitemShut
  {NoStop}%
\bibitem [{\citenamefont {Williams}\ \emph {et~al.}(2020)\citenamefont
  {Williams} \emph {et~al.}}]{Williams:2020mvu}%
  \BibitemOpen
  \bibfield  {author} {\bibinfo {author} {\bibfnamefont {D.}~\bibnamefont
  {Williams}} \emph {et~al.} (\bibinfo {collaboration} {IceCube}),\ }\href
  {\doibase 10.1016/j.nima.2018.11.109} {\bibfield  {journal} {\bibinfo
  {journal} {Nucl. Instrum. Meth.}\ }\textbf {\bibinfo {volume} {A952}},\
  \bibinfo {pages} {161650} (\bibinfo {year} {2020})}\BibitemShut {NoStop}%
\bibitem [{\citenamefont {Allison}\ \emph {et~al.}(2012)\citenamefont {Allison}
  \emph {et~al.}}]{Allison:2011wk}%
  \BibitemOpen
  \bibfield  {author} {\bibinfo {author} {\bibfnamefont {P.}~\bibnamefont
  {Allison}} \emph {et~al.},\ }\href {\doibase
  10.1016/j.astropartphys.2011.11.010} {\bibfield  {journal} {\bibinfo
  {journal} {Astropart. Phys.}\ }\textbf {\bibinfo {volume} {35}},\ \bibinfo
  {pages} {457} (\bibinfo {year} {2012})},\ \Eprint
  {http://arxiv.org/abs/1105.2854} {arXiv:1105.2854 [astro-ph.IM]} \BibitemShut
  {NoStop}%
\bibitem [{\citenamefont {Allison}\ \emph {et~al.}(2016)\citenamefont {Allison}
  \emph {et~al.}}]{Allison:2015eky}%
  \BibitemOpen
  \bibfield  {author} {\bibinfo {author} {\bibfnamefont {P.}~\bibnamefont
  {Allison}} \emph {et~al.} (\bibinfo {collaboration} {ARA}),\ }\href {\doibase
  10.1103/PhysRevD.93.082003} {\bibfield  {journal} {\bibinfo  {journal} {Phys.
  Rev.}\ }\textbf {\bibinfo {volume} {D93}},\ \bibinfo {pages} {082003}
  (\bibinfo {year} {2016})},\ \Eprint {http://arxiv.org/abs/1507.08991}
  {arXiv:1507.08991 [astro-ph.HE]} \BibitemShut {NoStop}%
\bibitem [{\citenamefont {Barwick}\ \emph
  {et~al.}(2015{\natexlab{a}})\citenamefont {Barwick} \emph
  {et~al.}}]{Barwick:2014rca}%
  \BibitemOpen
  \bibfield  {author} {\bibinfo {author} {\bibfnamefont {S.~W.}\ \bibnamefont
  {Barwick}} \emph {et~al.},\ }\href {\doibase 10.1109/TNS.2015.2468182}
  {\bibfield  {journal} {\bibinfo  {journal} {IEEE Trans. Nucl. Sci.}\ }\textbf
  {\bibinfo {volume} {62}},\ \bibinfo {pages} {2202} (\bibinfo {year}
  {2015}{\natexlab{a}})},\ \Eprint {http://arxiv.org/abs/1410.7369}
  {arXiv:1410.7369 [astro-ph.IM]} \BibitemShut {NoStop}%
\bibitem [{\citenamefont {Barwick}\ \emph
  {et~al.}(2015{\natexlab{b}})\citenamefont {Barwick} \emph
  {et~al.}}]{Barwick:2014pca}%
  \BibitemOpen
  \bibfield  {author} {\bibinfo {author} {\bibfnamefont {S.~W.}\ \bibnamefont
  {Barwick}} \emph {et~al.} (\bibinfo {collaboration} {ARIANNA}),\ }\href
  {\doibase 10.1016/j.astropartphys.2015.04.002} {\bibfield  {journal}
  {\bibinfo  {journal} {Astropart. Phys.}\ }\textbf {\bibinfo {volume} {70}},\
  \bibinfo {pages} {12} (\bibinfo {year} {2015}{\natexlab{b}})},\ \Eprint
  {http://arxiv.org/abs/1410.7352} {arXiv:1410.7352 [astro-ph.HE]} \BibitemShut
  {NoStop}%
\bibitem [{\citenamefont {Aartsen}\ \emph {et~al.}(2019)\citenamefont {Aartsen}
  \emph {et~al.}}]{Aartsen:2019swn}%
  \BibitemOpen
  \bibfield  {author} {\bibinfo {author} {\bibfnamefont {M.~G.}\ \bibnamefont
  {Aartsen}} \emph {et~al.} (\bibinfo {collaboration} {IceCube}),\ }\href@noop
  {} {\  (\bibinfo {year} {2019})},\ \Eprint {http://arxiv.org/abs/1911.02561}
  {arXiv:1911.02561 [astro-ph.HE]} \BibitemShut {NoStop}%
\bibitem [{\citenamefont {Adrian-Martinez}\ \emph {et~al.}(2016)\citenamefont
  {Adrian-Martinez} \emph {et~al.}}]{Adrian-Martinez:2016fdl}%
  \BibitemOpen
  \bibfield  {author} {\bibinfo {author} {\bibfnamefont {S.}~\bibnamefont
  {Adrian-Martinez}} \emph {et~al.} (\bibinfo {collaboration} {KM3Net}),\
  }\href {\doibase 10.1088/0954-3899/43/8/084001} {\bibfield  {journal}
  {\bibinfo  {journal} {J. Phys.}\ }\textbf {\bibinfo {volume} {G43}},\
  \bibinfo {pages} {084001} (\bibinfo {year} {2016})},\ \Eprint
  {http://arxiv.org/abs/1601.07459} {arXiv:1601.07459 [astro-ph.IM]}
  \BibitemShut {NoStop}%
\bibitem [{\citenamefont {Aloisio}\ \emph {et~al.}(2018)\citenamefont
  {Aloisio}, \citenamefont {Blasi}, \citenamefont {De~Mitri},\ and\
  \citenamefont {Petrera}}]{Aloisio:2017ooo}%
  \BibitemOpen
  \bibfield  {author} {\bibinfo {author} {\bibfnamefont {R.}~\bibnamefont
  {Aloisio}}, \bibinfo {author} {\bibfnamefont {P.}~\bibnamefont {Blasi}},
  \bibinfo {author} {\bibfnamefont {I.}~\bibnamefont {De~Mitri}}, \ and\
  \bibinfo {author} {\bibfnamefont {S.}~\bibnamefont {Petrera}},\ }\enquote
  {\bibinfo {title} {{Selected Topics in Cosmic Ray Physics}},}\ \ (\bibinfo
  {year} {2018})\ pp.\ \bibinfo {pages} {1--95},\ \Eprint
  {http://arxiv.org/abs/1707.06147} {arXiv:1707.06147 [astro-ph.HE]}
  \BibitemShut {NoStop}%
\bibitem [{\citenamefont {Chu}(1983)}]{Chu:83}%
  \BibitemOpen
  \bibfield  {author} {\bibinfo {author} {\bibfnamefont {W.~P.}\ \bibnamefont
  {Chu}},\ }\href {\doibase 10.1364/AO.22.000721} {\bibfield  {journal}
  {\bibinfo  {journal} {Appl. Opt.}\ }\textbf {\bibinfo {volume} {22}},\
  \bibinfo {pages} {721} (\bibinfo {year} {1983})}\BibitemShut {NoStop}%
\bibitem [{\citenamefont {Askaryan}(1962)}]{Askaryan:1962aa}%
  \BibitemOpen
  \bibfield  {author} {\bibinfo {author} {\bibfnamefont {G.~A.}\ \bibnamefont
  {Askaryan}},\ }\href@noop {} {\bibfield  {journal} {\bibinfo  {journal}
  {JETP}\ }\textbf {\bibinfo {volume} {14}},\ \bibinfo {pages} {441} (\bibinfo
  {year} {1962})},\ \bibinfo {note} {[Zh. Eksp. Teor.
  Fiz.41,616(1961)]}\BibitemShut {NoStop}%
\bibitem [{\citenamefont {Askaryan}(1989)}]{Askaryan:1989hk}%
  \BibitemOpen
  \bibfield  {author} {\bibinfo {author} {\bibfnamefont {G.~A.}\ \bibnamefont
  {Askaryan}},\ }\href@noop {} {\bibfield  {journal} {\bibinfo  {journal} {JETP
  Lett.}\ }\textbf {\bibinfo {volume} {50}},\ \bibinfo {pages} {478} (\bibinfo
  {year} {1989})},\ \bibinfo {note} {[Pisma Zh. Eksp. Teor.
  Fiz.50,446(1989)]}\BibitemShut {NoStop}%
\bibitem [{\citenamefont {Hoover}\ \emph {et~al.}(2010)\citenamefont {Hoover}
  \emph {et~al.}}]{Hoover:2010qt}%
  \BibitemOpen
  \bibfield  {author} {\bibinfo {author} {\bibfnamefont {S.}~\bibnamefont
  {Hoover}} \emph {et~al.} (\bibinfo {collaboration} {ANITA}),\ }\href
  {\doibase 10.1103/PhysRevLett.105.151101} {\bibfield  {journal} {\bibinfo
  {journal} {Phys. Rev. Lett.}\ }\textbf {\bibinfo {volume} {105}},\ \bibinfo
  {pages} {151101} (\bibinfo {year} {2010})},\ \Eprint
  {http://arxiv.org/abs/1005.0035} {arXiv:1005.0035 [astro-ph.HE]} \BibitemShut
  {NoStop}%
\bibitem [{\citenamefont {Schoorlemmer}\ \emph {et~al.}(2016)\citenamefont
  {Schoorlemmer} \emph {et~al.}}]{Schoorlemmer:2015afa}%
  \BibitemOpen
  \bibfield  {author} {\bibinfo {author} {\bibfnamefont {H.}~\bibnamefont
  {Schoorlemmer}} \emph {et~al.},\ }\href {\doibase
  10.1016/j.astropartphys.2016.01.001} {\bibfield  {journal} {\bibinfo
  {journal} {Astropart. Phys.}\ }\textbf {\bibinfo {volume} {77}},\ \bibinfo
  {pages} {32} (\bibinfo {year} {2016})},\ \Eprint
  {http://arxiv.org/abs/1506.05396} {arXiv:1506.05396 [astro-ph.HE]}
  \BibitemShut {NoStop}%
\bibitem [{\citenamefont {Gorham}\ \emph {et~al.}(2019)\citenamefont {Gorham}
  \emph {et~al.}}]{Gorham:2019guw}%
  \BibitemOpen
  \bibfield  {author} {\bibinfo {author} {\bibfnamefont {P.~W.}\ \bibnamefont
  {Gorham}} \emph {et~al.} (\bibinfo {collaboration} {ANITA}),\ }\href
  {\doibase 10.1103/PhysRevD.99.122001} {\bibfield  {journal} {\bibinfo
  {journal} {Phys. Rev.}\ }\textbf {\bibinfo {volume} {D99}},\ \bibinfo {pages}
  {122001} (\bibinfo {year} {2019})},\ \Eprint
  {http://arxiv.org/abs/1902.04005} {arXiv:1902.04005 [astro-ph.HE]}
  \BibitemShut {NoStop}%
\bibitem [{\citenamefont {Sissenwine}(1976)}]{USatmo:1976aaa}%
  \BibitemOpen
  \bibfield  {author} {\bibinfo {author} {\bibfnamefont {N.}~\bibnamefont
  {Sissenwine}},\ }\href {https://ntrs.nasa.gov/search.jsp?R=19770009539}
  {\bibfield  {journal} {\bibinfo  {journal} {NASA-TM-X-74335,
  NOAA-S/T-76-1562}\ } (\bibinfo {year} {1976})}\BibitemShut {NoStop}%
\bibitem [{\citenamefont {Neronov}\ \emph {et~al.}(2016)\citenamefont
  {Neronov}, \citenamefont {Semikoz}, \citenamefont {Vovk},\ and\ \citenamefont
  {Mirzoyan}}]{Neronov_2016}%
  \BibitemOpen
  \bibfield  {author} {\bibinfo {author} {\bibfnamefont {A.}~\bibnamefont
  {Neronov}}, \bibinfo {author} {\bibfnamefont {D.~V.}\ \bibnamefont
  {Semikoz}}, \bibinfo {author} {\bibfnamefont {I.}~\bibnamefont {Vovk}}, \
  and\ \bibinfo {author} {\bibfnamefont {R.}~\bibnamefont {Mirzoyan}},\ }\href
  {\doibase 10.1103/physrevd.94.123018} {\bibfield  {journal} {\bibinfo
  {journal} {Physical Review D}\ }\textbf {\bibinfo {volume} {94}} (\bibinfo
  {year} {2016}),\ 10.1103/physrevd.94.123018}\BibitemShut {NoStop}%
\bibitem [{\citenamefont {Kr\'olik}\ \emph {et~al.}(2019)\citenamefont
  {Kr\'olik}, \citenamefont {Djakonow}, \citenamefont {Plebaniak},
  \citenamefont {Przybylak}, \citenamefont {Szabelski},\ and\ \citenamefont
  {Wiencke}}]{krolik2019cherenkov}%
  \BibitemOpen
  \bibfield  {author} {\bibinfo {author} {\bibfnamefont {K.}~\bibnamefont
  {Kr\'olik}}, \bibinfo {author} {\bibfnamefont {A.}~\bibnamefont {Djakonow}},
  \bibinfo {author} {\bibfnamefont {Z.}~\bibnamefont {Plebaniak}}, \bibinfo
  {author} {\bibfnamefont {M.}~\bibnamefont {Przybylak}}, \bibinfo {author}
  {\bibfnamefont {J.}~\bibnamefont {Szabelski}}, \ and\ \bibinfo {author}
  {\bibfnamefont {L.}~\bibnamefont {Wiencke}},\ }\href@noop {} {\enquote
  {\bibinfo {title} {Cherenkov light from horizontal air shower},}\ } (\bibinfo
  {year} {2019}),\ \Eprint {http://arxiv.org/abs/1909.12007} {arXiv:1909.12007
  [astro-ph.IM]} \BibitemShut {NoStop}%
\bibitem [{\citenamefont {Bernlohr}(2008)}]{Bernlohr:2008mpk}%
  \BibitemOpen
  \bibfield  {author} {\bibinfo {author} {\bibfnamefont {K.}~\bibnamefont
  {Bernlohr}},\ }\href {\doibase 10.1016/j.astropartphys.2008.07.009}
  {\bibfield  {journal} {\bibinfo  {journal} {Astropart.\ Phys.}\ }\textbf
  {\bibinfo {volume} {30}},\ \bibinfo {pages} {149} (\bibinfo {year} {2008})},\
  \Eprint {http://arxiv.org/abs/0808.2253} {arXiv:0808.2253 [astro-ph.HE]}
  \BibitemShut {NoStop}%
\bibitem [{\citenamefont {Weast}(1986)}]{Weast:1986uoy}%
  \BibitemOpen
  \bibfield  {author} {\bibinfo {author} {\bibfnamefont {R.~C.}\ \bibnamefont
  {Weast}},\ }\href@noop {} {\emph {\bibinfo {title} {{CRC handbook of
  chemistry and physics}}}}\ (\bibinfo  {publisher} {67th Edition, CRC Press},\
  \bibinfo {year} {1986})\BibitemShut {NoStop}%
\bibitem [{\citenamefont {Ciddor}(1996)}]{Ciddor:1996aaa}%
  \BibitemOpen
  \bibfield  {author} {\bibinfo {author} {\bibfnamefont {P.~E.}\ \bibnamefont
  {Ciddor}},\ }\href {\doibase 10.1364/AO.35.001566} {\bibfield  {journal}
  {\bibinfo  {journal} {Appl. Opt.}\ }\textbf {\bibinfo {volume} {35}},\
  \bibinfo {pages} {1566} (\bibinfo {year} {1996})}\BibitemShut {NoStop}%
\bibitem [{\citenamefont {Hillas}(1982{\natexlab{a}})}]{Hillas:1982vn}%
  \BibitemOpen
  \bibfield  {author} {\bibinfo {author} {\bibfnamefont {A.}~\bibnamefont
  {Hillas}},\ }\href {\doibase 10.1088/0305-4616/8/10/016} {\bibfield
  {journal} {\bibinfo  {journal} {J. Phys. G}\ }\textbf {\bibinfo {volume}
  {8}},\ \bibinfo {pages} {1461} (\bibinfo {year}
  {1982}{\natexlab{a}})}\BibitemShut {NoStop}%
\bibitem [{\citenamefont {Hillas}(1982{\natexlab{b}})}]{Hillas:1982wz}%
  \BibitemOpen
  \bibfield  {author} {\bibinfo {author} {\bibfnamefont {A.}~\bibnamefont
  {Hillas}},\ }\href {\doibase 10.1088/0305-4616/8/10/017} {\bibfield
  {journal} {\bibinfo  {journal} {J.\ Phys.\ G}\ }\textbf {\bibinfo {volume}
  {8}},\ \bibinfo {pages} {1475} (\bibinfo {year}
  {1982}{\natexlab{b}})}\BibitemShut {NoStop}%
\bibitem [{\citenamefont {Elterman}(1968)}]{Elterman:1968}%
  \BibitemOpen
  \bibfield  {author} {\bibinfo {author} {\bibfnamefont {L.}~\bibnamefont
  {Elterman}},\ }\href@noop {} {}\bibinfo {number} {Tech. Rep. AFCRL-68-0153}\
  (\bibinfo {year} {1968})\BibitemShut {NoStop}%
\bibitem [{\citenamefont {Blitzstein}\ \emph {et~al.}(1970)\citenamefont
  {Blitzstein}, \citenamefont {Fliegel},\ and\ \citenamefont
  {Kondo}}]{Blitzstein:1970}%
  \BibitemOpen
  \bibfield  {author} {\bibinfo {author} {\bibfnamefont {W.}~\bibnamefont
  {Blitzstein}}, \bibinfo {author} {\bibfnamefont {H.~F.}\ \bibnamefont
  {Fliegel}}, \ and\ \bibinfo {author} {\bibfnamefont {Y.}~\bibnamefont
  {Kondo}},\ }\href {\doibase 10.1364/AO.9.002539} {\bibfield  {journal}
  {\bibinfo  {journal} {Appl. Opt.}\ }\textbf {\bibinfo {volume} {9}},\
  \bibinfo {pages} {2539} (\bibinfo {year} {1970})}\BibitemShut {NoStop}%
\bibitem [{\citenamefont {Serdyuchenko}\ \emph {et~al.}(2014)\citenamefont
  {Serdyuchenko}, \citenamefont {Gorshelev}, \citenamefont {Weber},
  \citenamefont {Chehade},\ and\ \citenamefont {Burrows}}]{AMT}%
  \BibitemOpen
  \bibfield  {author} {\bibinfo {author} {\bibfnamefont {A.}~\bibnamefont
  {Serdyuchenko}}, \bibinfo {author} {\bibfnamefont {V.}~\bibnamefont
  {Gorshelev}}, \bibinfo {author} {\bibfnamefont {M.}~\bibnamefont {Weber}},
  \bibinfo {author} {\bibfnamefont {W.}~\bibnamefont {Chehade}}, \ and\
  \bibinfo {author} {\bibfnamefont {J.~P.}\ \bibnamefont {Burrows}},\ }\href
  {\doibase 10.5194/amt-7-625-2014} {\bibfield  {journal} {\bibinfo  {journal}
  {Atmospheric Measurement Techniques}\ }\textbf {\bibinfo {volume} {7}},\
  \bibinfo {pages} {625} (\bibinfo {year} {2014})}\BibitemShut {NoStop}%
\bibitem [{\citenamefont {Gaisser}\ \emph {et~al.}(2016)\citenamefont
  {Gaisser}, \citenamefont {Engel},\ and\ \citenamefont
  {Resconi}}]{Gaisser:2016uoy}%
  \BibitemOpen
  \bibfield  {author} {\bibinfo {author} {\bibfnamefont {T.~K.}\ \bibnamefont
  {Gaisser}}, \bibinfo {author} {\bibfnamefont {R.}~\bibnamefont {Engel}}, \
  and\ \bibinfo {author} {\bibfnamefont {E.}~\bibnamefont {Resconi}},\
  }\href@noop {} {\emph {\bibinfo {title} {{Cosmic Rays and Particle Physics}:
  {2nd Edition}}}}\ (\bibinfo  {publisher} {Cambridge University Press},\
  \bibinfo {year} {2016})\BibitemShut {NoStop}%
\bibitem [{\citenamefont {Finlay}\ \emph {et~al.}(2010)\citenamefont {Finlay}
  \emph {et~al.}}]{GeoField}%
  \BibitemOpen
  \bibfield  {author} {\bibinfo {author} {\bibfnamefont {C.~C.}\ \bibnamefont
  {Finlay}} \emph {et~al.},\ }\href {\doibase 10.1111/j.1365-246X.2010.04804.x}
  {\bibfield  {journal} {\bibinfo  {journal} {Geophysical Journal
  International}\ }\textbf {\bibinfo {volume} {183}},\ \bibinfo {pages} {1216}
  (\bibinfo {year} {2010})},\ \Eprint
  {http://arxiv.org/abs/https://academic.oup.com/gji/article-pdf/183/3/1216/1785065/183-3-1216.pdf}
  {https://academic.oup.com/gji/article-pdf/183/3/1216/1785065/183-3-1216.pdf}
  \BibitemShut {NoStop}%
\bibitem [{\citenamefont {Homola}\ \emph {et~al.}(2015)\citenamefont {Homola},
  \citenamefont {Engel},\ and\ \citenamefont {Wilczy\'nski}}]{Homola:2014sra}%
  \BibitemOpen
  \bibfield  {author} {\bibinfo {author} {\bibfnamefont {P.}~\bibnamefont
  {Homola}}, \bibinfo {author} {\bibfnamefont {R.}~\bibnamefont {Engel}}, \
  and\ \bibinfo {author} {\bibfnamefont {H.}~\bibnamefont {Wilczy\'nski}},\
  }\href {\doibase 10.1016/j.astropartphys.2014.12.009} {\bibfield  {journal}
  {\bibinfo  {journal} {Astropart. Phys.}\ }\textbf {\bibinfo {volume} {60}},\
  \bibinfo {pages} {47} (\bibinfo {year} {2015})},\ \bibinfo {note} {[Erratum:
  Astropart.Phys. 65, 111 (2014)]},\ \Eprint {http://arxiv.org/abs/1405.4671}
  {arXiv:1405.4671 [astro-ph.IM]} \BibitemShut {NoStop}%
\bibitem [{\citenamefont {Bartoli}\ \emph {et~al.}(2014)\citenamefont {Bartoli}
  \emph {et~al.}}]{Bartoli:2014sza}%
  \BibitemOpen
  \bibfield  {author} {\bibinfo {author} {\bibfnamefont {B.}~\bibnamefont
  {Bartoli}} \emph {et~al.},\ }\href {\doibase 10.1103/PhysRevD.89.052005}
  {\bibfield  {journal} {\bibinfo  {journal} {Phys. Rev. D}\ }\textbf {\bibinfo
  {volume} {89}},\ \bibinfo {pages} {052005} (\bibinfo {year}
  {2014})}\BibitemShut {NoStop}%
\bibitem [{\citenamefont {Lafebre}\ \emph {et~al.}(2009)\citenamefont
  {Lafebre}, \citenamefont {Engel}, \citenamefont {Falcke}, \citenamefont
  {Horandel}, \citenamefont {Huege}, \citenamefont {Kuijpers},\ and\
  \citenamefont {Ulrich}}]{Lafebre:2009en}%
  \BibitemOpen
  \bibfield  {author} {\bibinfo {author} {\bibfnamefont {S.}~\bibnamefont
  {Lafebre}}, \bibinfo {author} {\bibfnamefont {R.}~\bibnamefont {Engel}},
  \bibinfo {author} {\bibfnamefont {H.}~\bibnamefont {Falcke}}, \bibinfo
  {author} {\bibfnamefont {J.}~\bibnamefont {Horandel}}, \bibinfo {author}
  {\bibfnamefont {T.}~\bibnamefont {Huege}}, \bibinfo {author} {\bibfnamefont
  {J.}~\bibnamefont {Kuijpers}}, \ and\ \bibinfo {author} {\bibfnamefont
  {R.}~\bibnamefont {Ulrich}},\ }\href {\doibase
  10.1016/j.astropartphys.2009.02.002} {\bibfield  {journal} {\bibinfo
  {journal} {Astropart.\ Phys.}\ }\textbf {\bibinfo {volume} {31}},\ \bibinfo
  {pages} {243} (\bibinfo {year} {2009})},\ \Eprint
  {http://arxiv.org/abs/0902.0548} {arXiv:0902.0548 [astro-ph.HE]} \BibitemShut
  {NoStop}%
\bibitem [{\citenamefont {Otte}(2019)}]{Otte:2018uxj}%
  \BibitemOpen
  \bibfield  {author} {\bibinfo {author} {\bibfnamefont {A.~N.}\ \bibnamefont
  {Otte}},\ }\href {\doibase 10.1103/PhysRevD.99.083012} {\bibfield  {journal}
  {\bibinfo  {journal} {Phys. Rev.}\ }\textbf {\bibinfo {volume} {D99}},\
  \bibinfo {pages} {083012} (\bibinfo {year} {2019})},\ \Eprint
  {http://arxiv.org/abs/1811.09287} {arXiv:1811.09287 [astro-ph.IM]}
  \BibitemShut {NoStop}%
\bibitem [{\citenamefont {Cillis}\ \emph {et~al.}(1999)\citenamefont {Cillis},
  \citenamefont {Fanchiotti}, \citenamefont {García~Canal},\ and\
  \citenamefont {Sciutto}}]{Cillis_1999}%
  \BibitemOpen
  \bibfield  {author} {\bibinfo {author} {\bibfnamefont {A.~N.}\ \bibnamefont
  {Cillis}}, \bibinfo {author} {\bibfnamefont {H.}~\bibnamefont {Fanchiotti}},
  \bibinfo {author} {\bibfnamefont {C.~A.}\ \bibnamefont {García~Canal}}, \
  and\ \bibinfo {author} {\bibfnamefont {S.~J.}\ \bibnamefont {Sciutto}},\
  }\href {\doibase 10.1103/physrevd.59.113012} {\bibfield  {journal} {\bibinfo
  {journal} {Physical Review D}\ }\textbf {\bibinfo {volume} {59}} (\bibinfo
  {year} {1999}),\ 10.1103/physrevd.59.113012}\BibitemShut {NoStop}%
\bibitem [{\citenamefont {Cummings}\ \emph {et~al.}(2019)\citenamefont
  {Cummings}, \citenamefont {Aloisio}, \citenamefont {Bertaina}, \citenamefont
  {Bisconti}, \citenamefont {Fenu},\ and\ \citenamefont
  {Salamida}}]{cummings2019complete}%
  \BibitemOpen
  \bibfield  {author} {\bibinfo {author} {\bibfnamefont {A.}~\bibnamefont
  {Cummings}}, \bibinfo {author} {\bibfnamefont {R.}~\bibnamefont {Aloisio}},
  \bibinfo {author} {\bibfnamefont {M.}~\bibnamefont {Bertaina}}, \bibinfo
  {author} {\bibfnamefont {F.}~\bibnamefont {Bisconti}}, \bibinfo {author}
  {\bibfnamefont {F.}~\bibnamefont {Fenu}}, \ and\ \bibinfo {author}
  {\bibfnamefont {F.}~\bibnamefont {Salamida}},\ }\href@noop {} {\enquote
  {\bibinfo {title} {A more complete phenomenology of tau lepton induced air
  showers},}\ } (\bibinfo {year} {2019}),\ \Eprint
  {http://arxiv.org/abs/1910.01021} {arXiv:1910.01021 [hep-ph]} \BibitemShut
  {NoStop}%
\bibitem [{\citenamefont {Crannell}\ and\ \citenamefont
  {Ormes}(1971)}]{CRANNELL1971179}%
  \BibitemOpen
  \bibfield  {author} {\bibinfo {author} {\bibfnamefont {C.}~\bibnamefont
  {Crannell}}\ and\ \bibinfo {author} {\bibfnamefont {J.}~\bibnamefont
  {Ormes}},\ }\href {\doibase https://doi.org/10.1016/0029-554X(71)90357-0}
  {\bibfield  {journal} {\bibinfo  {journal} {Nuclear Instruments and Methods}\
  }\textbf {\bibinfo {volume} {94}},\ \bibinfo {pages} {179 } (\bibinfo {year}
  {1971})}\BibitemShut {NoStop}%
\bibitem [{\citenamefont {Sullivan}(1971)}]{SULLIVAN19715}%
  \BibitemOpen
  \bibfield  {author} {\bibinfo {author} {\bibfnamefont {J.}~\bibnamefont
  {Sullivan}},\ }\href {\doibase https://doi.org/10.1016/0029-554X(71)90033-4}
  {\bibfield  {journal} {\bibinfo  {journal} {Nuclear Instruments and Methods}\
  }\textbf {\bibinfo {volume} {95}},\ \bibinfo {pages} {5} (\bibinfo {year}
  {1971})}\BibitemShut {NoStop}%
\bibitem [{\citenamefont {Venters}\ \emph {et~al.}(2019)\citenamefont
  {Venters}, \citenamefont {Reno}, \citenamefont {Krizmanic}, \citenamefont
  {Anchordoqui}, \citenamefont {Guépin},\ and\ \citenamefont
  {Olinto}}]{Venters:2019xwi}%
  \BibitemOpen
  \bibfield  {author} {\bibinfo {author} {\bibfnamefont {T.~M.}\ \bibnamefont
  {Venters}}, \bibinfo {author} {\bibfnamefont {M.~H.}\ \bibnamefont {Reno}},
  \bibinfo {author} {\bibfnamefont {J.~F.}\ \bibnamefont {Krizmanic}}, \bibinfo
  {author} {\bibfnamefont {L.~A.}\ \bibnamefont {Anchordoqui}}, \bibinfo
  {author} {\bibfnamefont {C.}~\bibnamefont {Guépin}}, \ and\ \bibinfo
  {author} {\bibfnamefont {A.~V.}\ \bibnamefont {Olinto}},\ }\href@noop {} {\
  (\bibinfo {year} {2019})},\ \Eprint {http://arxiv.org/abs/1906.07209}
  {arXiv:1906.07209 [astro-ph.HE]} \BibitemShut {NoStop}%
\bibitem [{\citenamefont {Budnev}\ \emph {et~al.}(2014)\citenamefont {Budnev}
  \emph {et~al.}}]{Tunka1}%
  \BibitemOpen
  \bibfield  {author} {\bibinfo {author} {\bibfnamefont {N.}~\bibnamefont
  {Budnev}} \emph {et~al.},\ }\href {\doibase 10.1088/1748-0221/9/09/C09021}
  {\bibfield  {journal} {\bibinfo  {journal} {Journal of Instrumentation}\
  }\textbf {\bibinfo {volume} {9}},\ \bibinfo {pages} {C09021} (\bibinfo {year}
  {2014})}\BibitemShut {NoStop}%
\bibitem [{\citenamefont {Berezhnev}\ \emph {et~al.}(2011)\citenamefont
  {Berezhnev} \emph {et~al.}}]{Tunka2}%
  \BibitemOpen
  \bibfield  {author} {\bibinfo {author} {\bibfnamefont {S.}~\bibnamefont
  {Berezhnev}} \emph {et~al.},\ }\href {\doibase 10.7529/ICRC2011/V01/0184}
  {\bibfield  {journal} {\bibinfo  {journal} {PoS}\ }\textbf {\bibinfo {volume}
  {ICRC2011}},\ \bibinfo {pages} {197} (\bibinfo {year} {2011})}\BibitemShut
  {NoStop}%
\bibitem [{\citenamefont {Krizmanic}\ \emph
  {et~al.}(2013{\natexlab{a}})\citenamefont {Krizmanic}, \citenamefont
  {Bergman},\ and\ \citenamefont {Sokolsky}}]{krizmanic2013nonimaging}%
  \BibitemOpen
  \bibfield  {author} {\bibinfo {author} {\bibfnamefont {J.}~\bibnamefont
  {Krizmanic}}, \bibinfo {author} {\bibfnamefont {D.}~\bibnamefont {Bergman}},
  \ and\ \bibinfo {author} {\bibfnamefont {P.}~\bibnamefont {Sokolsky}},\
  }\href@noop {} {\bibfield  {journal} {\bibinfo  {journal} {PoS}\ }\textbf
  {\bibinfo {volume} {ICRC2013}},\ \bibinfo {pages} {0365} (\bibinfo {year}
  {2013}{\natexlab{a}})},\ \Eprint {http://arxiv.org/abs/1307.3912}
  {arXiv:1307.3912 [astro-ph.IM]} \BibitemShut {NoStop}%
\bibitem [{\citenamefont {Krizmanic}\ \emph
  {et~al.}(2013{\natexlab{b}})\citenamefont {Krizmanic}, \citenamefont
  {Bergman},\ and\ \citenamefont {Sokolsky}}]{krizmanic2013modeling}%
  \BibitemOpen
  \bibfield  {author} {\bibinfo {author} {\bibfnamefont {J.}~\bibnamefont
  {Krizmanic}}, \bibinfo {author} {\bibfnamefont {D.}~\bibnamefont {Bergman}},
  \ and\ \bibinfo {author} {\bibfnamefont {P.}~\bibnamefont {Sokolsky}},\
  }\href@noop {} {\bibfield  {journal} {\bibinfo  {journal} {PoS}\ }\textbf
  {\bibinfo {volume} {ICRC2013}},\ \bibinfo {pages} {0366} (\bibinfo {year}
  {2013}{\natexlab{b}})},\ \Eprint {http://arxiv.org/abs/1307.3918}
  {arXiv:1307.3918 [astro-ph.IM]} \BibitemShut {NoStop}%
\bibitem [{\citenamefont {{Bergman}}\ \emph {et~al.}(2015)\citenamefont
  {{Bergman}}, \citenamefont {{Krizmanic}},\ and\ \citenamefont
  {{Tsunesada}}}]{2015ICRC...34..635B}%
  \BibitemOpen
  \bibfield  {author} {\bibinfo {author} {\bibfnamefont {D.}~\bibnamefont
  {{Bergman}}}, \bibinfo {author} {\bibfnamefont {J.}~\bibnamefont
  {{Krizmanic}}}, \ and\ \bibinfo {author} {\bibfnamefont {Y.}~\bibnamefont
  {{Tsunesada}}},\ }\href@noop {} {\bibfield  {journal} {\bibinfo  {journal}
  {PoS}\ }\textbf {\bibinfo {volume} {ICRC2015}},\ \bibinfo {eid} {635}
  (\bibinfo {year} {2015})}\BibitemShut {NoStop}%
\bibitem [{\citenamefont {{Krizmanic}}\ \emph {et~al.}(2015)\citenamefont
  {{Krizmanic}}, \citenamefont {{Bergman}},\ and\ \citenamefont
  {{Tsunesada}}}]{2015ICRC...34..562K}%
  \BibitemOpen
  \bibfield  {author} {\bibinfo {author} {\bibfnamefont {J.}~\bibnamefont
  {{Krizmanic}}}, \bibinfo {author} {\bibfnamefont {D.}~\bibnamefont
  {{Bergman}}}, \ and\ \bibinfo {author} {\bibfnamefont {Y.}~\bibnamefont
  {{Tsunesada}}},\ }\href@noop {} {\bibfield  {journal} {\bibinfo  {journal}
  {PoS}\ }\textbf {\bibinfo {volume} {ICRC2015}},\ \bibinfo {eid} {562}
  (\bibinfo {year} {2015})}\BibitemShut {NoStop}%
\bibitem [{\citenamefont {{Bergman}}\ \emph {et~al.}(2017)\citenamefont
  {{Bergman}}, \citenamefont {{Tsunesada}}, \citenamefont {{Krizmanic}},\ and\
  \citenamefont {{Omura}}}]{2017ICRC...35..415B}%
  \BibitemOpen
  \bibfield  {author} {\bibinfo {author} {\bibfnamefont {D.}~\bibnamefont
  {{Bergman}}}, \bibinfo {author} {\bibfnamefont {Y.}~\bibnamefont
  {{Tsunesada}}}, \bibinfo {author} {\bibfnamefont {J.~F.}\ \bibnamefont
  {{Krizmanic}}}, \ and\ \bibinfo {author} {\bibfnamefont {Y.}~\bibnamefont
  {{Omura}}},\ }\href@noop {} {\bibfield  {journal} {\bibinfo  {journal} {PoS}\
  }\textbf {\bibinfo {volume} {ICRC2017}},\ \bibinfo {eid} {415} (\bibinfo
  {year} {2017})}\BibitemShut {NoStop}%
\bibitem [{\citenamefont {Omura}\ \emph {et~al.}(2020)\citenamefont {Omura},
  \citenamefont {Nakai}, \citenamefont {Tsunesada}, \citenamefont {R~Bergman},\
  and\ \citenamefont {F~Krizmanic}}]{Omura:2020dzl}%
  \BibitemOpen
  \bibfield  {author} {\bibinfo {author} {\bibfnamefont {Y.}~\bibnamefont
  {Omura}}, \bibinfo {author} {\bibfnamefont {K.}~\bibnamefont {Nakai}},
  \bibinfo {author} {\bibfnamefont {Y.}~\bibnamefont {Tsunesada}}, \bibinfo
  {author} {\bibfnamefont {D.}~\bibnamefont {R~Bergman}}, \ and\ \bibinfo
  {author} {\bibfnamefont {J.}~\bibnamefont {F~Krizmanic}},\ }\href {\doibase
  10.22323/1.358.0379} {\bibfield  {journal} {\bibinfo  {journal} {PoS}\
  }\textbf {\bibinfo {volume} {ICRC2019}},\ \bibinfo {pages} {379} (\bibinfo
  {year} {2020})}\BibitemShut {NoStop}%
\bibitem [{\citenamefont {Bergman}\ \emph {et~al.}(2019)\citenamefont
  {Bergman}, \citenamefont {Krizmanic}, \citenamefont {Nakai}, \citenamefont
  {Omura},\ and\ \citenamefont {Tsunesada}}]{Bergman:2019f0}%
  \BibitemOpen
  \bibfield  {author} {\bibinfo {author} {\bibfnamefont {D.}~\bibnamefont
  {Bergman}}, \bibinfo {author} {\bibfnamefont {J.~F.}\ \bibnamefont
  {Krizmanic}}, \bibinfo {author} {\bibfnamefont {K.}~\bibnamefont {Nakai}},
  \bibinfo {author} {\bibfnamefont {Y.}~\bibnamefont {Omura}}, \ and\ \bibinfo
  {author} {\bibfnamefont {Y.}~\bibnamefont {Tsunesada}},\ }\href {\doibase
  10.22323/1.358.0189} {\bibfield  {journal} {\bibinfo  {journal} {PoS}\
  }\textbf {\bibinfo {volume} {ICRC2019}},\ \bibinfo {pages} {189} (\bibinfo
  {year} {2019})}\BibitemShut {NoStop}%
\end{thebibliography}%

\end{document}